%% file: ResourceTheoryCoherence.tex
\documentclass[a4paper,onecolumn,superscriptaddress,11pt,usenames,dvipsnames,svgnames,accepted=2020-09-01]{quantumarticle}
\usepackage{xcolor}
\definecolor{quantumviolet}{RGB}{79, 4, 134}
\definecolor{quantumgray}{RGB}{134,79,4}
\usepackage{enumitem}

\input{preamble.tex}
\newcommand{\PT}{\ensuremath \mathcal{P}}

\usepackage[UKenglish]{babel}

\newcommand{\decA}{\mathds{A}_{\resizebox{!}{1.5mm}{%
\InputIfFileExists{Diagrams/projWhite.tikz}{}{\input{./figures/Diagrams/projWhite.tikz}}}}}
\newcommand{\decAb}{\mathds{A}_{\resizebox{!}{1.5mm}{%
\InputIfFileExists{Diagrams/projBlack.tikz}{}{\input{./figures/Diagrams/projBlack.tikz}}}}}
\newcommand{\decB}{\mathds{B}_{\resizebox{!}{1.5mm}{%
\InputIfFileExists{Diagrams/projBlack.tikz}{}{\input{./figures/Diagrams/projBlack.tikz}}}}}

\begin{document}
\title{Compositional resource theories\\ of coherence}
\date{\today}
\author{John H. Selby}
\email{john.h.selby@gmail.com}
\affiliation{International Centre for Theory of Quantum Technologies, University of Gda\'nsk, Wita Stwosza 63, 80-308 Gda\'nsk, Poland}
\affiliation{Perimeter Institute for Theoretical Physics, 31 Caroline Street North, Waterloo, Ontario, N2L 2Y5, Canada}

\author{Ciar\'an M. Lee}
\email{ciaran.lee@ucl.ac.uk}
\affiliation{Department of Physics and Astronomy, University College London, Gower Street, WC1E 6BT, UK}
\begin{abstract}
Quantum coherence is one of the most important resources in quantum information theory. Indeed, preventing the loss of coherence is one of the most important technical challenges obstructing the development of large-scale quantum computers. Recently, there has been substantial progress in developing mathematical resource theories of coherence, paving the way towards its quantification and control. To date however, these resource theories have only been mathematically formalised within the realms of convex-geometry, information theory, and linear algebra. This approach is limited in scope, and makes it difficult to generalise beyond resource theories of coherence for single system quantum states.  In this paper we take a complementary perspective, showing that resource theories of coherence can instead be defined purely compositionally, that is, working with the mathematics of process theories, string diagrams and category theory. This new perspective offers several advantages: i) it unifies various existing approaches to the study of coherence, {for example, subsuming both speakable and unspeakable coherence}; ii) it provides a general treatment of the {compositional multi-system} setting; iii) it generalises immediately to the case of quantum channels, measurements, instruments, {and beyond} rather than just states; {iv) it can easily be generalised to the setting where there are multiple distinct sources of decoherence;} and, iv) it directly extends to arbitrary process theories, for example, generalised probabilistic theories and Spekkens toy model---providing the ability to operationally characterise coherence rather than relying on specific mathematical features of quantum theory for its description. More importantly, by providing a new, complementary, perspective on the resource of coherence, this work opens the door to the development of novel tools which would not be accessible from the linear algebraic mind set.
\end{abstract}
\maketitle

\newpage 

\section{Introduction}

The understanding and manipulation of coherence within quantum systems is central to the development of quantum technologies. More than this though, deeply understanding coherence is central to our understanding of quantum theory itself. Indeed, whilst loss of coherence in a quantum system is a critical problem in the development of quantum computers, it is essential for the emergent classical world in which we live our day to day lives \cite{zurek2003decoherence,zurek2009quantum,joos2013decoherence}. Developing better mathematical tools for addressing quantum coherence is therefore pertinent to both the foundations of physics as well as quantum engineering.

There has been great progress recently in the field of resource theories. Both the abstract mathematical formalism \cite{horodecki2013quantumness,brandao2015reversible,coecke2016mathematical,fritz2017resource,chitambar2019quantum} as well as important concrete examples have been developed and applied to a broad range of problems \cite{wolfe2019bell,schmid2019quantifying,streltsov2017colloquium,marvian2016quantify,chitambar2019quantum,sparaciari2017resource,ebler2018enhanced,rosset2018resource,lostaglio2015quantum}. The key benefit of the resource-theoretic perspective on quantum theory is that it takes qualitative ideas such as entanglement \cite{horodecki2009quantum,plenio2014introduction}, asymmetry \cite{ahmadi2013wigner}, coherence \cite{streltsov2017colloquium}, nonlocality \cite{wolfe2019bell} and so on \cite{del2015resource}, and turns them into quantitative studies. With such quantification comes the ability to optimise use of these resources and so efficiently manipulate them for the sake of performing practical tasks. In its infancy the study of resource theories was rather piecemeal. Recently however, a more systematic study has begun and unified frameworks \cite{coecke2016mathematical,fritz2017resource} have been developed for studying the abstract structure of resource theories. This has the added benefit that ideas developed in one resource theory can in a straightforward manner be ported over to another---we are no longer continually reinventing the wheel.

One of the key developments on the abstract formalism of resource theories has been the work of \cite{coecke2016mathematical}. This utilises the mathematics of category theory and process theories \cite{coecke2017picturing,selby2018reconstructing} to study resource theories in quantum theory and beyond. A key benefit of this approach is that it is intrinsically and explicitly compositional in nature. In particular, the mathematics of category theory comes with a natural diagrammatic representation which makes reasoning intuitive and greatly simplifies many complex calculations. For instance, this diagrammatic reasoning has been used to provide a natural environment for describing measurement-based quantum computation \cite{coecke2011interacting}, quantum teleportation \cite{coecke2006kindergarten}, and quantum secret sharing \cite{coecke2012strong} among many others \cite{coecke2017picturing}.

The study of resource theories of coherence has, to date, largely taken place from a more convex-geometric perspective, utilising tools such as majorization \cite{marshall2011b,nielsen1999conditions} and Schur-convex functions \cite{marshall2011b,gour2015resource}. For example see the works of \cite{winter2016operational,aberg2006quantifying,baumgratz2014quantifying,levi2014quantitative,yadin2016quantum,chitambar2016comparison}. However, this approach is somewhat limited in scope, and makes it difficult to generalise beyond resource theories of coherence for single system quantum states. Nowhere is understanding the manner in which {multisystem,} channel, and measurement coherence manifests as a resource more important than in a large-scale quantum computer, where multipartite systems evolve through the application of channels to eventually be measured. Hence understanding how coherence in such settings can be controlled and quantified---that is, used as a resource---is vital for achieving the promise of large-scale quantum computing.

In this work we take a complementary perspective and work entirely process-theoretically, equivalently we work entirely categorically, that is to say, diagrammatically. This has the benefit that it unifies many existing approaches. More importantly, it provides a general treatment of resources theories of coherence in a {compositional multi-system or multi-partite} setting, and also generalises immediately to the study of coherence of measurements, channels, general quantum instruments {and beyond}. {Moreover, the formalism can also be developed to allow for multiple distinct sources of decoherence for the same system.} Finally, it naturally extends to the study of coherence within alternative physical theories \cite{gogioso2017fantastic,gogioso2017categorical}, such as generalised probabilistic theories \cite{hardy2001quantum,barrett2007information,hardy2011reformulating,chiribella2011informational} and epistemically restricted theories \cite{spekkens2007evidence,coecke2011phase,backens2016complete,spekkens2016quasi}. This last point is useful as it provides the ability to characterise coherence in operational and physical terms, rather than via specific features of the mathematical formalism of quantum theory---such as Hilbert spaces and complex numbers. There is some existing work on studying resource theories in generalised theories such as \cite{krumm2017thermodynamics,chiribella2015entanglement,chiribella2017microcanonical,takagi2019general,barnum2015entropy} and this general approach has also been of use to deepen our understanding of computation \cite{lee2016information,krumm2019quantum,barrett2019computational,garner2018interferometric,barnum2018oracles,lee2016deriving,lee2016generalised,lee2015computation,lee2016bounds}, cryptography \cite{sikora2018simple,selby2018make,lami2018ultimate,sikora2019impossibility,barnum2011information,barnum2008nonclassicality,lee2018towards,lee2019device} and much more.

More importantly, however, than the application to specific alternative theories, is that this general approach demonstrates that the basic essence of coherence can be captured in terms of composition of processes, and hence, that the information-theoretic approach should not be taken to be primitive.

The remainder of this paper will be structured as follows.  In section \ref{Sec:Decoherence} we introduce a process-theoretic way of understanding decoherence. In section \ref{Sec:RTofCoherence} we introduce various resource theories of coherence which we can formulate from this compositional perspective. In this section we in particular: develop a resource theory for multi-partite scenarios; provide a unified treatment of speakable and unspeakable coherence; and generalise to the case in which there are multiple possible sources of decoherence.  Finally we discuss these results in section \ref{Sec:Discussion} and point out a wide range of topics for future work. Before this however we will provide some basic background material on process theories (sec.~\ref{Sec:ProcessTheories}) and resource theories (sec.~\ref{Sec:ResourceTheories} and app.~\ref{App:ResourceTheories}) which will be essential for understanding the rest of the paper.

\subsection{Process theories}\label{Sec:ProcessTheories}

We will provide a brief introduction to process theories here, for a more detailed introduction see, for example, \cite[Section 2]{selby2017process} or \cite{coecke2015categorical,coecke2017picturing,selby2018reconstructing}.

A process theory, $\mathcal{P}$, is defined by a collection of \emph{processes}, e.g.:
\beq
\InputIfFileExists{Diagrams/process.tikz}{}{\input{./figures/Diagrams/process.tikz}} \ \ \in \ \mathcal{P}
\eeq
is a process with \emph{input systems} $A$ and $C$ and \emph{output systems} $B$, $B$ and $A$ \footnote{{The labels, $A$, $B$, $C$, ..., on the systems are often called the system \emph{type}.}}. We interpret these systems as representing physical systems or degrees of freedom, the processes then represent how these physical systems evolve or are transformed. The defining feature of a process theory is that such processes can be wired together to make \emph{diagrams}, e.g.:
 \beq
\InputIfFileExists{Diagrams/diagram.tikz}{}{\input{./figures/Diagrams/diagram.tikz}}
 \eeq
which, itself must correspond to a process in the theory, in this case with input $A$ and output $D$. Note that it is only the \emph{connectivity} of these diagrams that matters, that is, how processes are connected together and the ordering of the inputs and outputs, rather than the precise position of the processes on the page. {Processes with no inputs, such as $a$ in the above diagram, are called \emph{states} and those with no outputs, such as $c$ in the above diagram, are called effects, those with neither inputs nor outputs are known as \emph{scalars}.}

Mathematically, these correspond to strict symmetric monoidal categories, however, whilst we could phrase our results in the language of category theory, it is more convenient to use their diagrammatic representation as string diagrams as this will suffice for our purposes.

In this paper we will consider only process theories which come with a way to discard systems, that is:
for each system $A$, there is a discarding map, denoted
\beq
\InputIfFileExists{Diagrams/discardA.tikz}{}{\input{./figures/Diagrams/discardA.tikz}}
\eeq
which respect the compositional structure, that is
\beq\label{Eq:DiscardComposition}
\InputIfFileExists{Diagrams/discardAB.tikz}{}{\input{./figures/Diagrams/discardAB.tikz}}\ \  =\ \  %
\InputIfFileExists{Diagrams/discardA.tikz}{}{\input{./figures/Diagrams/discardA.tikz}} %
\InputIfFileExists{Diagrams/discardB.tikz}{}{\input{./figures/Diagrams/discardB.tikz}}
\eeq
This condition states that discarding the components of a composite system is the same as discarding the composite system itself. We will say that a process, $f$, is \emph{causal} if it satisfies:
\beq\label{Eq:Causal}
\InputIfFileExists{Diagrams/causal.tikz}{}{\input{./figures/Diagrams/causal.tikz}}\ \ = \ \ %
\InputIfFileExists{Diagrams/discardA.tikz}{}{\input{./figures/Diagrams/discardA.tikz}}
\eeq
One can view the above as a way of stipulating that a process is deterministic. That is, a causal process implements a particular transformation on a given state that does not require post-selecting on a certain measurement outcome.

Many of our results can be extended to process theories without discarding, however discarding is very natural to assume for a physical theory so we will not concern ourselves with this mathematical generalisation here. We leave this as an exercise for the interested reader.

In many process theories we are also interested in measuring some property of a system, or, in controlling which process occurs in some experiment. To do so we can introduce classical systems to a process theory. If they are an output of some process then we view them as representing the classical system into which the outcome of the measurement is encoded (i.e., the position of some pointer). Diagrammatically we denote an $n$-outcome (non-destructive) measurement as:
\beq %
\InputIfFileExists{Diagrams/measurement.tikz}{}{\input{./figures/Diagrams/measurement.tikz}}\eeq
If they are the input then we can view them as some system which controls which process occurs (i.e., the position of some dial), we diagrammatically represent this as:
\beq %
\InputIfFileExists{Diagrams/controlled.tikz}{}{\input{./figures/Diagrams/controlled.tikz}}\eeq
Moreover, we can also have both classical inputs and classical outputs, in which case the input controls which (non-destructive) measurement is occurring, a.k.a. an instrument, which we denote as:
\beq %
\InputIfFileExists{Diagrams/controlledMeasurement.tikz}{}{\input{./figures/Diagrams/controlledMeasurement.tikz}}\eeq

\begin{example}[Quantum theory]The key example of interest for us here is quantum theory, see \cite{coecke2017picturing} for a detailed presentation of quantum theory as a process theory. In short however, systems in quantum theory correspond to Hilbert spaces $\mathcal{H}, \mathcal{K},...$ which, for simplicity, we take to be finite dimensional. Processes with input $\mathcal{H}$ and output $\mathcal{K}$ correspond to completely positive trace non-increasing (CPTNI) maps from $\mathcal{B[H]}$ to $\mathcal{B[K]}$. If there is no input (or output) system then we take the processes to correspond to be CPTNI maps from (or to) $\mathcal{B}[\mathds{C}]$, in particular this means that processes with no input correspond to (subnormalised) density matrices, those with no outputs to POVM elements and those with neither to probabilities. Composite systems are represented by the tensor product of the Hilbert spaces $\mathcal{HK}=\mathcal{H}\otimes\mathcal{K}$. The discarding map is given by the trace. Hence, in quantum theory the causal processes are those that are trace-preserving.\end{example}

\section{Resource theories}\label{Sec:ResourceTheories}

Here we provide a brief introduction to the process-theoretic approach to resource theories, {as originally introduced in} Ref.~\cite{coecke2016mathematical}. {A more detailed introduction to this framework is presented in App.~\ref{App:ResourceTheories}.}

{A resource theory specifies a set of resources and how they can be composed and freely converted. This idea, once appropriately formalised, is actually mathematically equivalent to the notion of a process theory that we have just introduced,} however, rather than interpreting systems as being physical systems and processes as their evolution, we instead interpret systems as being resources, and processes as ways in which we can freely transform {them. A free resource is then one that can be constructed by some free transformation `out of nothing'.}

{In Ref.~\cite{coecke2016mathematical} the authors show that there is, in some sense, a more fundamental structure which underpins a resource theory, which is known as a \emph{partitioned process theory}.} Specifically, given a process theory $\mathcal{P}$, in order to construct resource theories we must identify a sub-process theory of free processes $\mathcal{P}^{\mathsf{free}}$ -- the idea being that these are the physical processes which we can freely implement with no cost. For example, in constructing the resource theory of entanglement we may identify the LOCC operations as being free. This subset of free processes should be closed under composition, that is, if we can freely implement $f$ and freely implement $g$ then we should be able to freely do a composite of $f$ and $g$. Hence, rather than being an arbitrary subset of the processes, this subset must be a sub process theory. Such a pair $(\mathcal{P},\mathcal{P}^{\mathsf{free}})$ is known as a \emph{partitioned process theory}.

Given a partitioned process theory $(\mathcal{P},\mathcal{P}^{\mathsf{free}})$ Ref.~\cite{coecke2016mathematical} shows that there are various different resource theories, $\mathcal{R}$, which can be constructed. Roughly speaking, this consists of identifying the sorts of processes within $\mathcal{P}$ which you want to consider to be resources. For example, do we want to think of just the states in $\mathcal{P}$ as being resources, or do we want to think of the channels as being resources? There are various constructions of this sort which are discussed in Ref.~\cite{coecke2016mathematical}, {and we will formally introduce in the} App.~\ref{App:ResourceTheories}. Note, however, that the precise construction that should be used to define the resource theory will depend on exactly what one chooses to designate as a resource. In this sense, the partitioned process theory is really the more fundamental structure, and the particular resource theory that one constructs is situation dependent.

In the study of resource theories we are often not interested in the details of how we transform one resource into another (as is described by $\mathcal{R}$), but simply in whether or not there exists some way to freely transform one resource into another. That is, we often work simply with the {\emph{resource preorder}, $R$, denoted:
\beq
s \succ r
\eeq
for certain pairs of resources $s,r \in R$. This preorder $\succ$ means that $s$ can be freely converted into $r$, and hence, that $s$ is a more valuable resource than $r$. This resource preorder can be immediately obtained from a given resource theory simply by ignoring some of its structure.}

{It is therefore} important to see that, once the partitioned process theory has been defined, the rest is completely mechanical: once we know what resources we're interested in then we can construct the relevant resource theory and so we have (in principle at least) a characterisation of this preorder. The challenge for defining interesting resource theories in practice, therefore, boils down to defining interesting partitioned process theories. Hence, this is going to be our main focus in this paper.

\section{Decoherence} \label{Sec:Decoherence}

To understand coherence process-theoretically we will first introduce a process-theoretic understanding of decoherence. This will be crucial for our development of resource theories of coherence in section~\ref{Sec:RTofCoherence}. This formalism has been developed for both quantum and more general theories in the following papers \cite{coecke2017two,lee2017no,richens2017entanglement,selby2017leaks,gogioso2017categorical,heunen2013completely,selinger2008idempotents,scandolo2018possible,hefford2020hyper}, we will again not go into the full details of this but instead just present the fragment necessary for this paper.

Decoherence is generally an extremely complicated physical process involving an interaction of our system of interest $A$ with some uncontrolled environmental system, $E$, which we then lose access to. However, whilst the underlying mechanism is complex, the effective evolution of the system of interest is much simpler. Indeed, the essence of decoherence can be captured by the following simple definition.
\begin{definition}[Decoherence process]\label{Def:DecoherenceProcess}
A decoherence process for a system $A$, denoted:
\beq
\InputIfFileExists{Diagrams/idempotent.tikz}{}{\input{./figures/Diagrams/idempotent.tikz}}
\eeq
is a causal process, i.e.:
\beq
\InputIfFileExists{Diagrams/causalDecoherence.tikz}{}{\input{./figures/Diagrams/causalDecoherence.tikz}}\ \ = \ \ %
\InputIfFileExists{Diagrams/discardA.tikz}{}{\input{./figures/Diagrams/discardA.tikz}}
\eeq
which is idempotent, i.e.:
\beq
\InputIfFileExists{Diagrams/idempotentDef.tikz}{}{\input{./figures/Diagrams/idempotentDef.tikz}}\ \ = \ \ %
\InputIfFileExists{Diagrams/idempotent.tikz}{}{\input{./figures/Diagrams/idempotent.tikz}}
\eeq
\end{definition}
 This definition is motivated by the fact that decoherence should be causal, in that it is something that happens deterministically to our system rather than as a particular outcome of some measurement. Secondly, decoherence should be idempotent, as once coherence has been lost decohering again should have no further impact on the system. In general there are many different decoherence processes for a given system. Suitably choosing the decoherence processes is therefore an important part of defining a resource theory of coherence. {An alternative perspective, as suggested by one of the referees, is that such processes can be viewed as the physically realisable resource destroying maps \cite{liu2017resource}.}

 {Note that there are always trivial decoherence processes for any process theory, namely the identity process:
 \beq%
\InputIfFileExists{Diagrams/identityA.tikz}{}{\input{./figures/Diagrams/identityA.tikz}} \eeq
 these are both idempotent and causal so constitute (trivial) decoherence processes. {Clearly then, the interesting resource theories are to be found by considering non-trivial decoherence maps.}}

\begin{example}[Decoherence processes in quantum theory]
Let us briefly return to quantum theory to illustrate this with a key example. Consider a qubit, then an example of a decoherence process is the completely dephasing map, $\mathcal{D}$:
\beq
\mathcal{D}(\rho_A) =\sum_{i=0}^1 \ketbra{i_A}{i_A}\rho_A \ketbra{i_A}{i_A}
\eeq
{however, as discussed in Ref.~\cite{marvian2016quantify},}
{even just for a qubit there are qualitatively different forms that decoherence could take. For example, strictly, by our definitions, the identity map is a decoherence process, albeit an extremely uninteresting one. Moreover, any discard-prepare channel is a valid decoherence process, again however one that will not lead to particularly interesting resource theories. In the end what can be shown \cite{coecke2017two} is that the qualitatively different forms of decoherence can be classified as the sub C*-algebras of the quantum system. Intuitively, these C*-algebras simply correspond to the different block-diagonal forms that can be embedded within the quantum system, for example, if we consider a qutrit then one C*-algebra is the set of $3\times3$ density matrices, one is given by density matrices $2\times 2$ block and a $1\times 1$ block, and the final one is given by the completely diagonal density matrices.} {Our notion of decoherence here is therefore substantially more general, even for quantum theory, than the traditionally studied case of the completely dephasing map (see, e.g., \cite{streltsov2017colloquium}).}
\end{example}
These decoherence processes are actually all that we need to define certain resource theories of coherence---in particular those representing ``unspeakable coherence'' as coined in \cite{marvian2016quantify}. However, for other resource theories of coherence we will need a little more information about the decoherence mechanisms that we are interested in. Specifically, this will be necessary for representing ``speakable coherence'' \cite{marvian2016quantify}.

In particular, we need to specify not just the decoherence process, but a closely related process that we will term the \emph{decoherence mechanism}\footnote{Previously known as preleaks in \cite{selby2017leaks} but this terminology is not so useful here.}. Intuitively, a decoherence mechanism specifies the resultant state of the environmental system in addition to the evolution of the system of interest---this is important to study how information about the system of interest leaks into the environment.

\begin{definition}[Decoherence mechanism]\label{Def:DecoherenceMechanism} A decoherence mechanism is a process denoted
\beq
\InputIfFileExists{Diagrams/preleak.tikz}{}{\input{./figures/Diagrams/preleak.tikz}}
\eeq
which induces a decoherence process on the system $A$ when the environment system $E$ is discarded, i.e.:
\beq
\InputIfFileExists{Diagrams/preleakDef.tikz}{}{\input{./figures/Diagrams/preleakDef.tikz}}\ \ = \ \ %
\InputIfFileExists{Diagrams/idempotent.tikz}{}{\input{./figures/Diagrams/idempotent.tikz}}
\eeq
\end{definition}
Note that given a particular decoherence process there will, in general, be many decoherence mechanisms which could induce it. For example, the decoherence process itself can actually be viewed as a decoherence mechanism, just one where the system $E$ is {the trivial `no system' system}. Part of the challenge of constructing interesting resource theories of coherence is therefore to pick suitably interesting decoherence mechanisms.

\begin{example}[Decoherence mechanisms in quantum theory]
To give one example, a decoherence mechanism, $\mathcal{B}$, that could induce the completely dephasing map, $\mathcal{D}$, could be the following map from a qubit $A$ to itself together with an environmental qubit $E$:
\beq
\mathcal{B}(\rho_A) = \sum_{i=0}^1\ketbra{i_Ai_E}{i_A}\rho_A \ketbra{i_A}{i_Ai_E}
\eeq
\end{example}

Typically decoherence is viewed as a process that happens to states, that is, given some state $s$ of system $A$, if we have some decoherence process for $A$ then we can define the decohered state as:
\beq%
\InputIfFileExists{Diagrams/DecoheredState.tikz}{}{\input{./figures/Diagrams/DecoheredState.tikz}}\eeq
and hence we can define the \emph{decohered state space} as the collection of all such states. However, unlike in many other approaches to studying decoherence we can also define decoherence of general processes, that is, given some process $f:A\to B$, and a decoherence mechanism for both $A$ and $B$ then we can define the decohered process as those that are pre and post-composed with decoherence processes:
\beq%
\InputIfFileExists{Diagrams/DecoheredProcessNEW.tikz}{}{\input{./figures/Diagrams/DecoheredProcessNEW.tikz}}\eeq
and so we can define the \emph{decohered process space} as the collection of all such processes. Due to idempotence of the decoherence processes, we can equivalently define the decohered processes as being those that satisfy:
\beq
\InputIfFileExists{Diagrams/simpleProcess.tikz}{}{\input{./figures/Diagrams/simpleProcess.tikz}} \ \ = \ \ %
\InputIfFileExists{Diagrams/DecoheredProcessNEW.tikz}{}{\input{./figures/Diagrams/DecoheredProcessNEW.tikz}}
\eeq
or, again equivalently, the pair of constraints that:
\beq
\InputIfFileExists{Diagrams/DecoheredProcessIN.tikz}{}{\input{./figures/Diagrams/DecoheredProcessIN.tikz}} \ \ = \ \ %
\InputIfFileExists{Diagrams/simpleProcess.tikz}{}{\input{./figures/Diagrams/simpleProcess.tikz}} \ \ = \ \ %
\InputIfFileExists{Diagrams/DecoheredProcessOUT.tikz}{}{\input{./figures/Diagrams/DecoheredProcessOUT.tikz}}
\eeq

Going further however, we can define decoherence for a process theory $\mathcal{P}$ itself. To do so, for each system $A$, we choose a decoherence process, we denote this collection as:
\beq
\mathsf{\mathsf{dec}} := \left\{%
\InputIfFileExists{Diagrams/idempotent.tikz}{}{\input{./figures/Diagrams/idempotent.tikz}}\ \middle|\  A \in \mathcal{P}\right\}
\eeq
Note that if we have classical systems in our process theory then we will usually want to interpret these as having trivial decoherence, that is, where the chosen decoherence process is simply the identity.

We can then define the set of decohered processes in $\mathcal{P}$, which we denote as $\mathcal{P}^{\mathsf{dec}}$, as:
\beq \label{eq:DecoheredSubTheory}
\mathcal{P}^{\mathsf{dec}}:=\left\{%
\InputIfFileExists{Diagrams/ProcessNEW.tikz}{}{\input{./figures/Diagrams/ProcessNEW.tikz}} \in \mathcal{P}\ \ \middle|\ \ %
\InputIfFileExists{Diagrams/ProcessNEW.tikz}{}{\input{./figures/Diagrams/ProcessNEW.tikz}}\ \ = \ \ %
\InputIfFileExists{Diagrams/DecoheredProcessNEW.tikz}{}{\input{./figures/Diagrams/DecoheredProcessNEW.tikz}}\right\}
\eeq
 This is the minimal structure that we require to construct resource theories. However, it will commonly be the case that we want $\mathcal{P}^{\mathsf{dec}}$ to itself be a process theory, that is, it should be closed under wiring together processes. Intuitively, if we compose two processes which don't possess any coherence, then the resulting process also shouldn't have any coherence. That is, the act of composition doesn't spontaneously create coherence. Additionally, such a feature would be desirable, for instance, if we want to view decoherence as describing how an entire theory emerges from another, for example, to describe how classical theory emerges from quantum theory via decoherence.

To ensure that $\mathcal{P}^{\mathsf{dec}}$ is closed under composition we must impose certain constraints on the choice of the decoherence processes. The specific constraint to impose will depend on the specific situation that one has in mind, however, there are two choices that seem to crop up naturally.

\begin{example}\label{ex:LocalDecoherenceComposition}
The easiest way to enforce such a compositionality constraint is to simply demand, for all systems $A$ and $B$, that:
\beq\label{eq:LocalDecoherence}
\InputIfFileExists{Diagrams/decoherenceComposition.tikz}{}{\input{./figures/Diagrams/decoherenceComposition.tikz}}
\eeq
This would be a suitable choice, for example, if we view these decoherences as occurring in a quantum computer where it is typically assumed that noise is uncorrelated.
\end{example}
\begin{lemma} \label{lem:23} If the decoherence process for  composite system is the composite of the decoherence processes for the components, then $\mathcal{P}^{\mathsf{dec}}$ is closed under composition.
\end{lemma}
\proof
See appendix \ref{app:23}
\endproof
 This constraint on decoherence for composite systems may often be the appropriate choice\footnote{{Indeed, it is by far the most commonly studied situation in the literature.}}, however, there are cases where this is too restrictive. For example, if we want to consider decoherence arising from loss of some global reference frame then we shouldn't lose the correlations between local systems. In this case then there is a set of weaker conditions which are satisfied which still ensure that $\mathcal{P}^{\mathsf{dec}}$ is closed under composition.
 \begin{example}\label{ex:GlobalDecoherenceComposition}
Another way to enforce compositionality of $\mathcal{P}^{\mathsf{dec}}$ is by demanding i) that the composition of decoherence processes are decohered processes, i.e. that for all systems $A$,$B$ and $C$ we have:
 \beq\label{eq:24}
\InputIfFileExists{Diagrams/decoherenceConstraintNEW2.tikz}{}{\input{./figures/Diagrams/decoherenceConstraintNEW2.tikz}} \ \ = \ \ %
\InputIfFileExists{Diagrams/decoherenceConstraintNEW3.tikz}{}{\input{./figures/Diagrams/decoherenceConstraintNEW3.tikz}}
\eeq
and, ii) that if we compose a global with a local decoherence we get a pair of local decoherences, i.e. that for all systems $A$ and $B$ we have:
\beq\label{eq:25}
\InputIfFileExists{Diagrams/decoherenceConstrantNEW4.tikz}{}{\input{./figures/Diagrams/decoherenceConstrantNEW4.tikz}}
\eeq
  \end{example}
 \begin{lemma}\label{lem:24}
 The conjunction of these two constraints (equations \ref{eq:24} and \ref{eq:25}) imply that $\mathcal{P}^{\mathsf{dec}}$ is closed under composition.
 \end{lemma}
 \proof
 See appendix \ref{app:24}
 \endproof

If one were to pick the decoherence processes such that $\mathcal{P}^{\mathsf{dec}}$ is indeed closed under composition, then this defines a partitioned process theory\footnote{{Strictly this is not a partitioned process theory in the sense of \cite{coecke2016mathematical} as the free set does not include identities. There are various ways around this technicality, for instance, we could simply add identity processes to the free set.}}. One might then be tempted to use this as the definition of a partitioned process theory from which to construct resource theories, that is, defining a partitioned process theory by $(\mathcal{P},\mathcal{P}^{\mathsf{free}}:=\mathcal{P}^{\mathsf{dec}})$\footnote{{Indeed such a definition has been studied in \cite{saxena2020dynamical}.}}. However, this is far too restrictive a definition of free processes. Specifically, we can see that the resource theory of states has an extremely boring preorder. To see this, note that if take the decohered processes as being the free processes then any free process, $f$, maps any arbitrary (potentially non-free) state $\psi$ in $\PT$ into the free (i.e. decohered) partition:
\beq
\forall \psi \in \mathcal{P}, f\in\mathcal{P}^{\mathsf{dec}}\qquad %
\InputIfFileExists{Diagrams/trivialOrder.tikz}{}{\input{./figures/Diagrams/trivialOrder.tikz}} \ \ \in \ \  \mathcal{P}^{\mathsf{dec}}
\eeq
Hence, the preordering of the resource theory of states for this partitioning is too coarse grained to capture interesting phenomena. We will therefore, in the following section, explore how to construct interesting resource theories of coherence.

\section{Resource theories of coherence}\label{Sec:RTofCoherence}

The above mathematical concepts that we have introduced provide us with a tool-box for constructing resource theories of coherence. The specific resource theory that one constructs from these will really depend on the specific application that one has in mind. In this section we will therefore illustrate this toolbox with some example resource theories, however, we stress that this is not intended to be exhaustive and should be viewed more as a pedagogical introduction to these tools.

In doing so, we encounter some of the resource theories previously discussed in the literature---such as Decoherence Invariant Operations ($\mathsf{DIO}$), and Translationally Invariant Operations ($\mathsf{TIO}$) \cite{streltsov2017colloquium}---as special cases. Moreover, we explicitly construct a resource theory of coherence for multipartite systems, the free processes of which we term Complete Decoherence Invariant Operations. Importantly, we show that the constraints on free processes in this theory are not implied by any previous theory in the literature.

There are (at least) two qualitative choices to be made when deciding which partitioned process theory to construct, firstly, there is the question of whether we want the coherence to be speakable or unspeakable---as defined in \cite{marvian2016quantify}. Intuitively, consider the two quantum states $\frac{1}{\sqrt{2}}(\ket{0}+\ket{1})$ and $\frac{1}{\sqrt{2}}(\ket{0}+\ket{2})$ of a qutrit. If these are treated as equivalent resources then we are dealing with \emph{speakable} coherence, where all that matters is whether there is a superposition, not which subspaces are involved in the superposition (such as would be the case in quantum computation). On the other hand, if these are inequivalent resources then we are dealing with \emph{unspeakable} coherence, where which subspaces appear in superposition is relevant (such as would be the case if we were using these states for metrology). Secondly, there is the question of how to deal with composite systems, namely, is coherence a local or a global phenomena\footnote{This depends on what we view as being the source of decoherence, in general within quantum computation we assume that errors are local, and so systems decohere locally. On the other hand, if we view decoherence as arising from the loss of some global reference frame then it makes sense to define decoherence as a global phenomena.}.  In both cases we can define ``intermediate'' cases as well but the two extremes are the most interesting. Then, when moving from the partitioned process theory to a resource theory there is the question of what is the resource, are they states, general processes, collections of processes or something more complicated such as quantum combs?

Regardless of all of these choices, we will demand that our resource theories must abide by one basic principle:
\begin{principle}\label{basicPrinciple}
\emph{Free processes must preserve the set of decohered processes.}
\end{principle}

Intuitively, free processes cannot create coherence spontaneously, and so should map any process lacking coherence to another. {We will show how to formalise this intuition in the following subsection.} Interestingly, {we will then see that} some free processes defined in the literature, such as Maximally Incoherent Operations ($\mathsf{MIO}$) \cite{streltsov2017colloquium}, fail to satisfy this principle.

\subsection{Single system speakable coherence}

We will start with a very simple example which will help to illustrate the basic concepts: a process theory with only a single system (and the trivial system) and, hence, {there will be just one relevant} decoherence map. From this, we will construct the resource theory of states. This is not a particularly physically well motivated process theory as any physical theory will have non-trivial composite systems, however, we can still see many of the important features appearing in this simplified setting.

Given a process theory $\mathcal{P}$ with a single system $A$ we can pick a particular decoherence process:
\beq
\InputIfFileExists{Diagrams/idempotent.tikz}{}{\input{./figures/Diagrams/idempotent.tikz}}
\eeq
 we can then, as before, define the decohered subtheory as the collection of processes:
\beq
\mathcal{P}^{\mathsf{dec}} := \left\{%
\InputIfFileExists{Diagrams/DecoheredState.tikz}{}{\input{./figures/Diagrams/DecoheredState.tikz}}\ \ ,\ \ %
\InputIfFileExists{Diagrams/DecoheredProcessNEW2.tikz}{}{\input{./figures/Diagrams/DecoheredProcessNEW2.tikz}}\ \ ,\ \ %
\InputIfFileExists{Diagrams/DecoheredEffectNEW.tikz}{}{\input{./figures/Diagrams/DecoheredEffectNEW.tikz}} \right\}
\eeq
Now, a minimal requirement for the free processes must be that they preserve the set of decohered processes. For example, a free process $g:A\to A$ must satisfy:
\beq
\InputIfFileExists{Diagrams/MinConstraint1.tikz}{}{\input{./figures/Diagrams/MinConstraint1.tikz}} \ \ = \ \ %
\InputIfFileExists{Diagrams/MinConstraint2.tikz}{}{\input{./figures/Diagrams/MinConstraint2.tikz}}
\eeq
that is, it must preserve the set of free states. This example is commonly taken as the only constraint on free processes, corresponding, at least in the quantum case, to the free processes known as MIO in the literature \cite{streltsov2017colloquium}. However, now that we are taking a process-theoretic viewpoint we can see that this is not a good choice for free operations, as this is just one way that $g$ could be combined with a decohered process. Another way that it can be combined with a decohered process is to compose it with a decohered effect, in which case we additionally obtain the constraint:
\beq
\InputIfFileExists{Diagrams/MinConstraint3.tikz}{}{\input{./figures/Diagrams/MinConstraint3.tikz}} \ \ = \ \ %
\InputIfFileExists{Diagrams/MinConstraint4.tikz}{}{\input{./figures/Diagrams/MinConstraint4.tikz}}
\eeq
Indeed, if we consider all such possibilities we can show that our free processes must satisfy:
\begin{lemma}That a process $g \in \mathcal{P}^{\mathsf{free}}$ preserves the decohered subtheory $\mathcal{P}^{\mathsf{dec}}$ is equivalent to the condition:
\beq%
\InputIfFileExists{Diagrams/speakableProcessNEW.tikz}{}{\input{./figures/Diagrams/speakableProcessNEW.tikz}}\eeq
\end{lemma}
\proof To see this it suffices to consider composition of a free process with the decohered process which is just the decoherence process itself. It is simple to see that the decoherence process is indeed in $\mathcal{P}^{\mathsf{dec}}$ due to idempotence of the decoherence process. Pre-composing $g$ with the decoherence process must give another process $h$ in the decohered theory,
\beq
\InputIfFileExists{Diagrams/preserveIdempotent.tikz}{}{\input{./figures/Diagrams/preserveIdempotent.tikz}}
\eeq
then, using idempotence of decoherence, it is simple to show that this gives us
\beq
\InputIfFileExists{Diagrams/preserveIdempotent2.tikz}{}{\input{./figures/Diagrams/preserveIdempotent2.tikz}}
\eeq
Similarly, we can ask that the post-composing $g$ with the decoherence map must give another process in the decohered theory which allows us to show that:
\beq
\InputIfFileExists{Diagrams/preserveIdempotent3.tikz}{}{\input{./figures/Diagrams/preserveIdempotent3.tikz}}
\eeq
the conjunction of these two gives the desired constraint.  What remains is to show that this implies that any decohered process is preserved by {$g$}.
This follows immediately from the above constraint together with eq.~\eqref{eq:DecoheredSubTheory} in the definition of the decohered subtheory.
\endproof
Such a constraint on the free processes is known as $\mathsf{DIO}$ in the literature \cite{streltsov2017colloquium}. That is, we have shown that
\beq
\mathcal{P}^{\mathsf{free}} \subseteq \mathsf{DIO}
\eeq
Therefore, any proposal for a resource theory of coherence which use a set of free processes larger than $\mathsf{DIO}$ are inconsistent with {our basic principle of free processes (Principle~\ref{basicPrinciple}),} namely, they map decohered processes to processes which are not decohered\footnote{{This is closely related to a result in \cite{gour2017quantum} in the context of resource destroying maps in quantum theory which similarly singles out $\mathsf{DIO}$ as the largest relevant set of free processes.}}.

If we work with the definition that $\mathcal{P}^{\mathsf{free}}:= \mathsf{DIO}$ then we can, using the methods discussed in the earlier sections, define from this a resource theory of coherence of states (or measurements, or channels), and from that obtain a preorder amongst the states (or measurements, or channels) which fully quantifies their coherence.

One should note at this point that one of the major differences of this approach to more conventional approaches to the study of coherence is the basic building blocks that we use. Typically, the starting point of the study of coherence is to pick a preferred basis, and from there to define coherence relative to that basis. This then demands that we explain why we chose that particular basis, what made it distinct from any other basis? Here we do not pick a basis at all.  Instead, we pick a particular decoherence map. This decoherence map is given to us by the physics of the situation which tells us how our system decoheres. We could, if we wanted, characterise the decoherence via process tomography, but this is not necessary to prove general results that would hold regardless of the precise nature of decoherence. Such a decoherence process may in turn pick out a preferred basis (for example, if it is a totally dephasing map in quantum theory), however, importantly this preferred basis is picked out by the physical decoherence process rather than being something that we had to choose at the start.

Let us now illustrate these definitions with a simple example.
\begin{example}[Resource theory of coherence of qubit states]
Consider a process theory describing a single qubit, with the decoherence map given by $\mathcal{D}(\_) =\sum_{i=0}^1 \ketbra{i}{i}\_ \ketbra{i}{i}$. The decohered processes are then: states are of the form $\sum_i p_i \ketbra{i}{i}$ where $p_i \in [0,1]$ and $\sum_i p_i =1$; effects are of the form $\sum_i r_i \ketbra{i}{i}$ where $r_i \in [0,1]$; and, general transformations are of the form $\sum_{ij} s_{ij} \ketbra{i}{j}\_ \ketbra{j}{i}$ where $s_{ij}$ is a stochastic matrix.

The free processes are then those that preserve the set of decohered processes, or equivalently, that commute with the dephasing map $\mathcal{D}$. That is: for states and effects these are simply the decohered states and effects; whilst for transformations they are those such that satisfy $\mathcal{D}\circ \mathcal{E} = \mathcal{E}\circ \mathcal{D}$.

We can then explicitly construct the relevant resource theory for states, {see Ex.~\ref{ex:RTofStates} in App.~\ref{App:ResourceTheories}.} Specifically, the resources correspond to arbitrary qubit states $\rho$, and there is a transformation from resource $\rho$ to resource $\sigma$ if and only if there is a free process mapping $\rho$ to $\sigma$:
\beq
\exists \ %
\InputIfFileExists{Diagrams/SimpleRTS.tikz}{}{\input{./figures/Diagrams/SimpleRTS.tikz}} \quad \iff \quad \exists \mathcal{E} \text{ s.t. } \mathcal{D}\circ \mathcal{E} = \mathcal{E}\circ \mathcal{D} \text{ and } %
\InputIfFileExists{Diagrams/SimpleRTS2.tikz}{}{\input{./figures/Diagrams/SimpleRTS2.tikz}} = %
\InputIfFileExists{Diagrams/SimpleRTS3.tikz}{}{\input{./figures/Diagrams/SimpleRTS3.tikz}}
\eeq
If within the resource theory there exists such a process $\mathcal{E}:\rho \to \sigma$ then we write that $\rho \succ \sigma$, or in words, that $\rho$ is at least as resourceful as $\sigma$.
\end{example}

The fact that free processes must be $\mathsf{DIO}$ does not, however, imply that taking a smaller set of free processes cannot be justified. There are two instances in which this may be well motivated. The first, which we will return to later, is when we investigate composite systems, the second, which we consider next, is when we deal with unspeakable coherence.

\subsection{Single system unspeakable coherence}
We can deal with unspeakable coherence within our toy example of a process theory with a single system. To do so, rather than just picking a particular decoherence process for the system, we must also pick a suitable decoherence mechanism, that is:
\beq
\InputIfFileExists{Diagrams/idempotent.tikz}{}{\input{./figures/Diagrams/idempotent.tikz}} \qquad \text{and} \qquad %
\InputIfFileExists{Diagrams/preleak.tikz}{}{\input{./figures/Diagrams/preleak.tikz}}
\eeq
such that
\beq
\InputIfFileExists{Diagrams/idempotent.tikz}{}{\input{./figures/Diagrams/idempotent.tikz}}\ \ = \ \ %
\InputIfFileExists{Diagrams/preleakDef.tikz}{}{\input{./figures/Diagrams/preleakDef.tikz}}
\eeq
the natural generalisation of the definition of the $\mathsf{DIO}$ processes is then given by the following condition:
\beq
\InputIfFileExists{Diagrams/unspeakableProcessNEW.tikz}{}{\input{./figures/Diagrams/unspeakableProcessNEW.tikz}}
\eeq
This condition states that not only do we require that free processes commute with decoherence, but also that the effect of decoherence on the environment is left invariant by the free processes.
This immediately implies the condition of $\mathsf{DIO}$ as can be seen by discarding system $E$ in the above equation:
\beq
\InputIfFileExists{Diagrams/unspeakableProcessNEWDiscard.tikz}{}{\input{./figures/Diagrams/unspeakableProcessNEWDiscard.tikz}}\ \  \implies\ \  %
\InputIfFileExists{Diagrams/speakableProcessNEW.tikz}{}{\input{./figures/Diagrams/speakableProcessNEW.tikz}}
\eeq
Therefore this set of processes is contained within $\mathsf{DIO}$. Precisely how restrictive this condition is will depend on the precise nature of this decoherence mechanism. It could be a trivial restriction, for example, in the case that
\beq
\InputIfFileExists{Diagrams/preleak.tikz}{}{\input{./figures/Diagrams/preleak.tikz}} \ \ = \ \ \ %
\InputIfFileExists{Diagrams/trivialPreLeak.tikz}{}{\input{./figures/Diagrams/trivialPreLeak.tikz}}
\eeq
for some normalised state $s$, this clearly reduces to the definition of $\mathsf{DIO}$. However, in other cases this is much more restrictive. For example, in quantum theory, if we take the decoherence mechanism to be
\beq
\mathcal{B}(\rho_A) = \sum_{i=0}^1\ketbra{i_Ai_E}{i_A}\rho_A \ketbra{i_A}{i_Ai_E}
\eeq
then, this corresponds to the set of processes known as $\mathsf{TIO}$ in the literature \cite{streltsov2017colloquium}.

To see this more generally, let us sketch out how to describe decoherence as arising from the lack of access to some reference frame {within quantum theory}\footnote{{This clearly applies much more broadly than just to quantum theory but would require more process-theoretic formalism than we introduce here. A formal treatment of the general case is therefore left to future work.}}.  Suppose we have some group $G$ which represents the possible orientations of the reference frame. Then we have some physical system which encodes this orientation in a set of perfectly distinguishable states:
\beq\left\{%
\InputIfFileExists{Diagrams/orthogStates.tikz}{}{\input{./figures/Diagrams/orthogStates.tikz}}\right\}_{\gamma\in G}\eeq
which are perfectly distinguished by the effects:
\beq\left\{%
\InputIfFileExists{Diagrams/orthogEffects.tikz}{}{\input{./figures/Diagrams/orthogEffects.tikz}}\right\}_{\gamma\in G}\eeq
Then, we introduce the action of the group on the physical system:
\beq\left\{%
\InputIfFileExists{Diagrams/groupAction.tikz}{}{\input{./figures/Diagrams/groupAction.tikz}}\right\}_{\gamma\in G}\eeq
We can then write the decoherence mechanism arising from this reference frame as:
\beq
\InputIfFileExists{Diagrams/referenceFrame.tikz}{}{\input{./figures/Diagrams/referenceFrame.tikz}}
\eeq
such that the decoherence process is given by loss of the reference frame as:
\beq
\InputIfFileExists{Diagrams/referenceFrameLoss.tikz}{}{\input{./figures/Diagrams/referenceFrameLoss.tikz}}
\eeq
{such a decoherence process is known as a $G$-twirling map in the  literature.}
Recall that our constraint on free processes is that:
\beq
\InputIfFileExists{Diagrams/unspeakableProcessNEW.tikz}{}{\input{./figures/Diagrams/unspeakableProcessNEW.tikz}}
\eeq
In the case we are considering this means that:
\beq
\InputIfFileExists{Diagrams/referenceFrameCommuting.tikz}{}{\input{./figures/Diagrams/referenceFrameCommuting.tikz}}
\eeq
which implies that
\beq
\forall \gamma'\in G \ \ \ %
\InputIfFileExists{Diagrams/referenceFrameCommuting2.tikz}{}{\input{./figures/Diagrams/referenceFrameCommuting2.tikz}}
\eeq
and so, due to distinguishability of the reference frame states, we find that
\beq
\forall \gamma'\in G \ \ \ \ %
\InputIfFileExists{Diagrams/referenceFrameCommuting3.tikz}{}{\input{./figures/Diagrams/referenceFrameCommuting3.tikz}}
\eeq
That is, $f$ commutes with the action of the group $G$ -- precisely the definition of $\mathsf{TIO}$. Note that by varying the distinguishability of the states $s_\gamma$ we can interpolate between this $\mathsf{TIO}$ and the $\mathsf{DIO}$ conditions. We no longer have a speakable~vs.~unspeakable dichotomy, but rather we have a spectrum between the two.

\subsection{Multi-system speakable coherence}

We now turn to much more interesting process theories, of which quantum theory is an example, where we have multiple different systems---we must now contend with the added difficulty (and interest!) afforded by parallel composition. This is the first time a resource theory of coherence has been developed for multiple systems, {however, the multipartite case has also been studied from the perspective of combining entanglement and coherence into a single resource theory \cite{chitambar2016relating,streltsov2017towards} using restricted classes of LOCC operations.}.

We will continue to adhere to our basic principle {(Principle~\ref{basicPrinciple})}: free processes must preserve the set of decohered processes. For example, if $h\in \mathcal{P}^{\mathsf{dec}}$ and $g\in\mathcal{P}^{\mathsf{free}}$, then it must be the case that:
\beq %
\InputIfFileExists{Diagrams/CompositionExample1.tikz}{}{\input{./figures/Diagrams/CompositionExample1.tikz}} \ \ \in \mathcal{P}^{\mathsf{dec}} \eeq
However, the sorts of composites that we consider here cannot be completely arbitrary, for example, we do not expect the parallel composite of a decohered process with a free process to itself be decohered. For instance the identity process belongs to the free set but we would expect, for any $h\in\mathcal{P}^{\mathsf{dec}}$, that:
\beq%
\InputIfFileExists{Diagrams/CompositionExample2.tikz}{}{\input{./figures/Diagrams/CompositionExample2.tikz}} \ \ \not\in \mathcal{P}^{\mathsf{dec}}\eeq
Our key principle should therefore be formalised as follows:
\begin{definition}\label{def:minConstraint}
The minimal constraint that we will impose on $\mathcal{P}^{\mathsf{free}}$ is therefore that, if $g$ is free then, for any $h,h' \in \mathcal{P}^{\mathsf{dec}}$ with suitable system types we have:
\beq %
\InputIfFileExists{Diagrams/CompositionExample1.tikz}{}{\input{./figures/Diagrams/CompositionExample1.tikz}} \ \ \in \mathcal{P}^{\mathsf{dec}} \quad  \text{and} \quad %
\InputIfFileExists{Diagrams/CompositionExample3.tikz}{}{\input{./figures/Diagrams/CompositionExample3.tikz}}\in \mathcal{P}^{\mathsf{dec}} \eeq
\end{definition}

It is simple to show  that this is equivalent to a ``complete'' notion of decoherence invariance, which we coin $\mathsf{cDIO}$.
\begin{definition}\label{def:CDIO}Given a set of decoherence processes $\mathsf{\mathsf{dec}}$ we an define the set of $\mathsf{cDIO}$ processes as:
\beq\mathsf{cDIO} := \left\{\begin{tikzpicture}
	\begin{pgfonlayer}{nodelayer}
		\node [style=none] (0) at (0, 1.25) {};
		\node [style={small box}] (1) at (0, -0) {$g$};
		\node [style={right label}] (2) at (0, 0.75) {$B$};
		\node [style=none] (3) at (0, -1.25) {};
		\node [style={right label}] (4) at (0, -1) {$A$};
	\end{pgfonlayer}
	\begin{pgfonlayer}{edgelayer}
		\draw [qWire] (1) to (0.center);
		\draw [qWire] (3.center) to (1);
	\end{pgfonlayer}
\end{tikzpicture}
 \in \mathcal{P}\ \middle|\ %
\InputIfFileExists{Diagrams/CDIODef1.tikz}{}{\input{./figures/Diagrams/CDIODef1.tikz}} =  \ \ %
\InputIfFileExists{Diagrams/CDIODef2.tikz}{}{\input{./figures/Diagrams/CDIODef2.tikz}}\ \  \forall C\in\mathcal{P}\right\}\eeq
where \beq%
\InputIfFileExists{Diagrams/wideDecoherence.tikz}{}{\input{./figures/Diagrams/wideDecoherence.tikz}} \in \mathsf{\mathsf{dec}} \eeq
\end{definition}

\begin{lemma}\label{lem:34}A process $g$ is in $\mathsf{cDIO}$ if and only if it satisfies the minimal constraint on free processes (def.~\ref{def:minConstraint}). That is, def.~\ref{def:minConstraint} and def.~\ref{def:CDIO} are equivalent.
\end{lemma}
\proof
See appendix \ref{app:34}.
\endproof

This means that the largest set of free processes that we can consider are the $\mathsf{cDIO}$ processes, i.e. $\mathcal{P}^{\mathsf{free}} \subseteq \mathsf{cDIO}$.

Now, if we are in the special case, {which has been most commonly studied in the literature, in which}
we have that global decoherence processes are just the composite of local decoherence processes, that is, as discussed in example \ref{ex:GlobalDecoherenceComposition}:
\beq%
\InputIfFileExists{Diagrams/decoherenceComposition.tikz}{}{\input{./figures/Diagrams/decoherenceComposition.tikz}}\eeq
then $\mathsf{cDIO}=\mathsf{DIO}$, i.e. we can conclude that a process is $\mathsf{cDIO}$ simply by checking that:
\beq%
\InputIfFileExists{Diagrams/speakableProcessNEW.tikz}{}{\input{./figures/Diagrams/speakableProcessNEW.tikz}}\eeq

One may therefore be interested in asking whether $\mathsf{cDIO}$ is a genuine restriction over $\mathsf{DIO}$, or whether they are always equivalent. That is, can we find a situation in which $\mathsf{cDIO}\subsetneq \mathsf{DIO}$?
This turns out to be the case---as we will prove via an explicit example shortly---hence, $\mathsf{cDIO}$ is a genuine restriction over $\mathsf{DIO}$ in certain circumstances.

To illustrate this, let us consider the case within quantum theory, where we are viewing decoherence as loss of a global reference system. That is, the decoherence process for a composite system is given by:
\beq%
\InputIfFileExists{Diagrams/globalRefFrameLoss.tikz}{}{\input{./figures/Diagrams/globalRefFrameLoss.tikz}} \ \ = \ \ %
\InputIfFileExists{Diagrams/globalRefFrameLoss2.tikz}{}{\input{./figures/Diagrams/globalRefFrameLoss2.tikz}}\eeq
It is simple to check that this satisfies the constraint on composition of decoherence processes described in example \ref{ex:GlobalDecoherenceComposition} and so that $\mathcal{P}^{\mathsf{dec}}$ is closed under composition. Intuitively, this form of decoherence loses any correlations to some external reference frame, but maintains correlations between subsystems. To see that this is not equivalent to $\mathsf{DIO}$ let us consider just a pair of qubit systems, $G$ to be the group $\mathds{Z}_2$ with representation $R_\gamma$ as either the identity, {$\mathds{1}$,} or a bit flip, {$X$}. Then, it is simple to compute, by considering the equation:
\beq%
\InputIfFileExists{Diagrams/CDIOExample1.tikz}{}{\input{./figures/Diagrams/CDIOExample1.tikz}} \ \ = \ \ %
\InputIfFileExists{Diagrams/CDIOExample2.tikz}{}{\input{./figures/Diagrams/CDIOExample2.tikz}}\eeq
that:
\beq%
\InputIfFileExists{Diagrams/CDIOExample3.tikz}{}{\input{./figures/Diagrams/CDIOExample3.tikz}} \ \ = \ \ %
\InputIfFileExists{Diagrams/CDIOExample4.tikz}{}{\input{./figures/Diagrams/CDIOExample4.tikz}}\eeq
Hence, that the $\mathsf{cDIO}$ transformations are those that are translationally invariant with respect to the action of $\mathds{Z}_2$. This is known to be a genuine restriction over $\mathsf{DIO}$, {see, for example, \cite{marvian2016quantify}.}

We have therefore found, by considering composite systems, that we are forced to revise our notions of what free processes are, even when those processes themselves do not involve composite systems. This, in retrospect is perhaps not surprising, as we know that complete positivity is a genuine restriction over positivity---the restriction from $\mathsf{DIO}$ to $\mathsf{cDIO}$ is analogous. It is therefore crucial that when constructing resource theories generally, and, in particular, for resource theories of coherence, that we do not just focus on single system phenomena but consider general forms of composition and the restrictions that they impose.

Finally, if we want to use the $\mathsf{cDIO}$ processes to define resource theories, that is, if we want to work with $\mathcal{P}^{\mathsf{free}} = \mathsf{cDIO}$ then it must be the case that this defines a partitioned process theory. In particular, we must check that $\mathsf{cDIO}$ processes are closed under composition.
\begin{lemma} $\mathsf{cDIO}$ processes are closed under composition. \end{lemma}
\proof Consider two processes in $\mathsf{cDIO}$, $f$ and $g$ and a general composite process:
\beq%
\InputIfFileExists{Diagrams/CDIOClosure1.tikz}{}{\input{./figures/Diagrams/CDIOClosure1.tikz}}\eeq
we want to show that this too is in $\mathsf{cDIO}$. The proof of this is extremely simple:
\beq%
\InputIfFileExists{Diagrams/CDIOClosure2.tikz}{}{\input{./figures/Diagrams/CDIOClosure2.tikz}} \ \ = \ \ %
\InputIfFileExists{Diagrams/CDIOClosure3.tikz}{}{\input{./figures/Diagrams/CDIOClosure3.tikz}} \ \ = \ \ %
\InputIfFileExists{Diagrams/CDIOClosure4.tikz}{}{\input{./figures/Diagrams/CDIOClosure4.tikz}} \eeq
where the first equality uses the fact that $f$ is in $\mathsf{cDIO}$ and the second that $g$ is in $\mathsf{cDIO}$.
\endproof

\subsection{Multi-system unspeakable coherence}

Given the definition of speakable coherence above for multiple systems and of unspeakable coherence for single systems, it is simple to see how one should define speakable coherence for multiple systems. Namely, for each system $A$ we define a decoherence mechanism:
\beq
\left\{%
\InputIfFileExists{Diagrams/preleak.tikz}{}{\input{./figures/Diagrams/preleak.tikz}}\middle| A\in\mathcal{P} \right\}
\eeq
Then, we can define the free processes as those that commute with decoherence mechanisms in a complete sense, that is, for all systems $C$ we demand:
\beq\label{eq:speakMulti}
\InputIfFileExists{Diagrams/SpeakCDIODef1.tikz}{}{\input{./figures/Diagrams/SpeakCDIODef1.tikz}} \ =  \ \ %
\InputIfFileExists{Diagrams/SpeakCDIODef2.tikz}{}{\input{./figures/Diagrams/SpeakCDIODef2.tikz}}
\eeq
Note that one limitation of this approach is that we require that each decoherence mechanism shares the same environment system $E$. One could avoid this at the expense of introducing encoding and decoding maps to transform between different environments. However, we can always avoid this by letting $E$ be suitably large such that any other environment can be trivially embedded within it. Moreover, this level of generality does not seem to add anything new conceptually so we leave its exploration to future works in which it may be relevant.

It is important to now check two things. The first check, that this set of free processes is contained within $\mathsf{cDIO}$ as $\mathsf{cDIO}$ was the minimal requirement that we wanted our free processes to satisfy. Specifically, when we define the $\mathsf{cDIO}$ processes relative to the decoherence processes:
\beq\mathsf{\mathsf{\mathsf{dec}}} : = \left\{ %
\InputIfFileExists{Diagrams/idempotent.tikz}{}{\input{./figures/Diagrams/idempotent.tikz}} \ \ :=\ \ %
\InputIfFileExists{Diagrams/preLeakDiscard.tikz}{}{\input{./figures/Diagrams/preLeakDiscard.tikz}}\middle| A\in\mathcal{P} \right\}\eeq
Secondly, we must check that processes satisfying this constraint are closed under composition such that they can be used to define a partitioned process theory. These two checks are indeed satisfied as is shown in the following simple results.
\begin{lemma} The processes that commute with decoherence mechanisms must also commute with the associated decoherence processes. \end{lemma}
\proof
The proof  is trivial, simply compose the `commuting with decoherence mechanisms' condition with a discarding process on the environment system:
\beq%
\InputIfFileExists{Diagrams/SpeakCDIODef1.tikz}{}{\input{./figures/Diagrams/SpeakCDIODef1.tikz}}=  \ \ %
\InputIfFileExists{Diagrams/SpeakCDIODef2.tikz}{}{\input{./figures/Diagrams/SpeakCDIODef2.tikz}}  \implies \ \ %
\InputIfFileExists{Diagrams/CTIOinCDIO1.tikz}{}{\input{./figures/Diagrams/CTIOinCDIO1.tikz}}  =  \ \ %
\InputIfFileExists{Diagrams/CTIOinCDIO2.tikz}{}{\input{./figures/Diagrams/CTIOinCDIO2.tikz}}\eeq
which immediately, given the definition of the decoherence processes, gives the desired result:
\beq%
\InputIfFileExists{Diagrams/CDIODef1.tikz}{}{\input{./figures/Diagrams/CDIODef1.tikz}} =  \ \ %
\InputIfFileExists{Diagrams/CDIODef2.tikz}{}{\input{./figures/Diagrams/CDIODef2.tikz}}\eeq
\endproof

\begin{lemma}
Processes satisfying \eqref{eq:speakMulti} are closed under composition.
\end{lemma}
\proof
Consider two processes, $f$ and $g$ satisfying this constraint and a general composite process:
\beq%
\InputIfFileExists{Diagrams/CDIOClosure1.tikz}{}{\input{./figures/Diagrams/CDIOClosure1.tikz}}\eeq
we want to show that this too satisfies this constraint. The proof of this is again extremely simple:
\beq%
\InputIfFileExists{Diagrams/CTIOClosure2.tikz}{}{\input{./figures/Diagrams/CTIOClosure2.tikz}} \ \ = \ \ %
\InputIfFileExists{Diagrams/CTIOClosure3.tikz}{}{\input{./figures/Diagrams/CTIOClosure3.tikz}} \ \ = \ \ %
\InputIfFileExists{Diagrams/CTIOClosure4.tikz}{}{\input{./figures/Diagrams/CTIOClosure4.tikz}} \eeq
where the first equality uses the fact that $f$ satisfies the constraint and the second that $g$ does as well.
\endproof
This therefore defines a valid partitioned process theory in which the free partitioned preserves the set of decohered processes. Hence we can, as in the other cases, use this to define resource theories of states, measurements, general processes, and so on.

Again, as we found in the single system case, we can `interpolate' between speakable and unspeakable coherence by modifying the precise details of the chosen decoherence mechanisms. Intuitively, if the environment perfectly encodes the eigenspace then we will have resource theories of unspeakable coherence, and if it encodes no information then it will be speakable coherence.

Like in the single-system case, a particular instantiation of this will correspond to translationally invariant operations. However, like in the multipartite speakable case, we will find that we need to make this a `complete' notion which we call $\mathsf{cTIO}$. Specifically, given some group $G$ and some representation on each composite system, then we define:
\beq
\mathsf{cTIO} := \left\{%
\InputIfFileExists{Diagrams/simpleProcess.tikz}{}{\input{./figures/Diagrams/simpleProcess.tikz}}\in\mathcal{P}\ \middle|\ %
\InputIfFileExists{Diagrams/DefcTIO.tikz}{}{\input{./figures/Diagrams/DefcTIO.tikz}}=%
\InputIfFileExists{Diagrams/DefcTIO2.tikz}{}{\input{./figures/Diagrams/DefcTIO2.tikz}} \forall C\in\mathcal{P}, \gamma\in G\right\}
\eeq
Again, like in the case of $\mathsf{cDIO}$, we find that $\mathsf{cTIO}$ can be a genuine restriction over $\mathsf{TIO}$. {However, if we are in the special case where the group representations factorise over composite systems then it will be the case that $\mathsf{cTIO}$ and $\mathsf{TIO}$ coincide.} Therefore, it is imperative to specify the group representation, not just for single systems, but for composite systems as well, in order to define consistent resource theories of unspeakable coherence.
{
\subsection{Multiple sources of decoherence}\label{sec:multidecoherence}
In all of the above we have considered just a single decoherence process/mechanism per system. There will be, however, situations in which we may be interested in multiple different decoherence processes/mechanisms for the same system. In this section we will briefly introduce a formalism that allows us to handle such situations. This is closely related to the categorical construction known as the Karoubi envelope/Cauchy completion/splitting idempotents which has been studied within the context of Categorical Quantum Mechanics in \cite{coecke2017two,heunen2013completely,selinger2008idempotents} and in more general theories in \cite{gogioso2017categorical,selby2017leaks,lee2017no,richens2017entanglement,hefford2020hyper}.

Intuitively, the Karoubi envelope $\mathcal{K}$ of a process theory, $\mathcal{P}$, can be thought of as the process theory which contains every possible decohered subtheory $\mathcal{P}^{\mathsf{dec}}$ in a minimal way. That is, for any choice of decoherence processes $\mathsf{dec}$, it is the case that $\mathcal{P}^{\mathsf{dec}}\subseteq \mathcal{K[P]}$. More formally it is defined as follows:

\begin{definition}[Karoubi envelope, $\mathcal{K}$]
Systems labeled in $\mathcal{K[P]}$ are labeled by arbitrary pairs:
\beq
\InputIfFileExists{Diagrams/Karoubi2.tikz}{}{\input{./figures/Diagrams/Karoubi2.tikz}}\resizebox{!}{3mm}{$\left(A,%
\InputIfFileExists{Diagrams/Karoubi1.tikz}{}{\input{./figures/Diagrams/Karoubi1.tikz}}\right)$}
\eeq
where $A$ is a system in $\mathcal{P}$ and $%
\InputIfFileExists{Diagrams/smallDecohere.tikz}{}{\input{./figures/Diagrams/smallDecohere.tikz}}$ is an idempotent on $A$. We introduce the shorthand notation for systems:
\beq
\decA:=\left(A,%
\InputIfFileExists{Diagrams/Karoubi1.tikz}{}{\input{./figures/Diagrams/Karoubi1.tikz}}\right)
\eeq
and, noting that there can be multiple different idempotents per system we will distinguish them by their colour, hence we can have systems such as $\decA$ and $\decAb$ which differ only by their associated idempotent.

We can then note that these compose via:
\beq\label{eq:KPComp}
\InputIfFileExists{Diagrams/Karoubi2.tikz}{}{\input{./figures/Diagrams/Karoubi2.tikz}}\resizebox{!}{3mm}{$\left(A,%
\InputIfFileExists{Diagrams/Karoubi1.tikz}{}{\input{./figures/Diagrams/Karoubi1.tikz}}\right)$}
\ \
\InputIfFileExists{Diagrams/Karoubi2.tikz}{}{\input{./figures/Diagrams/Karoubi2.tikz}}\resizebox{!}{3mm}{$\left(B,%
\InputIfFileExists{Diagrams/Karoubi3.tikz}{}{\input{./figures/Diagrams/Karoubi3.tikz}}\right)$}
\quad := \quad
\InputIfFileExists{Diagrams/Karoubi2.tikz}{}{\input{./figures/Diagrams/Karoubi2.tikz}}\resizebox{!}{3mm}{$\left(AB,%
\InputIfFileExists{Diagrams/Karoubi1.tikz}{}{\input{./figures/Diagrams/Karoubi1.tikz}}%
\InputIfFileExists{Diagrams/Karoubi3.tikz}{}{\input{./figures/Diagrams/Karoubi3.tikz}}\right)$}
\eeq
Processes in $\mathcal{K[P]}$ from  $\decA$  to $\decB$  are a subset of the processes in $\mathcal{P}$ from $A$ to $B$, namely, those satisfying:
\beq
\begin{tikzpicture}
	\begin{pgfonlayer}{nodelayer}
		\node [style=none] (0) at (0, 1.25) {};
		\node [style={small box}] (1) at (0, -0) {$f$};
		\node [style={right label}] (2) at (0, 0.75) {$B$};
		\node [style=none] (3) at (0, -1.25) {};
		\node [style={right label}] (4) at (0, -1) {$A$};
	\end{pgfonlayer}
	\begin{pgfonlayer}{edgelayer}
		\draw [qWire] (1) to (0);
		\draw [qWire] (3) to (1);
	\end{pgfonlayer}
\end{tikzpicture}
\quad =\quad
\begin{tikzpicture}
	\begin{pgfonlayer}{nodelayer}
		\node [style=proj,fill=black] (0) at (0, 1.25) {};
		\node [style={small box}] (1) at (0, -0) {$f$};
		\node [style=none] (2) at (0, 2) {};
		\node [style={right label}] (3) at (0, 1.75) {$B$};
		\node [style={right label}] (4) at (0, 0.75) {$B$};
		\node [style=none] (5) at (0, -2) {};
		\node [style={right label}] (6) at (0, -1.75) {$A$};
		\node [style=proj] (7) at (0, -1.25) {};
		\node [style={right label}] (8) at (0, -1) {$A$};
	\end{pgfonlayer}
	\begin{pgfonlayer}{edgelayer}
		\draw [qWire] (1) to (0);
		\draw [qWire] (0) to (2.center);
		\draw [qWire] (7) to (5.center);
		\draw [qWire] (7) to (1);
	\end{pgfonlayer}
\end{tikzpicture}
\eeq
Processes in $\mathcal{K[P]}$ then compose exactly as their associated processes would in $\mathcal{P}$.
\end{definition}

The connection to decoherence is clear -- the processes in $\mathcal{K[P]}$ from  $\decA$  to $\decB$ correspond to decohered processes where system $A$ has been decohered by $\begin{tikzpicture}
	\begin{pgfonlayer}{nodelayer}
		\node [style=proj] (0) at (0, -0) {};
		\node [style=none] (1) at (0, -0.5) {};
		\node [style=none] (2) at (0, 0.5) {};
	\end{pgfonlayer}
	\begin{pgfonlayer}{edgelayer}
		\draw [qWire] (2.center) to (0);
		\draw [qWire] (0) to (1.center);
	\end{pgfonlayer}
\end{tikzpicture}
$ and system $B$ by $\begin{tikzpicture}
	\begin{pgfonlayer}{nodelayer}
		\node [style=proj,fill=black] (0) at (0, -0) {};
		\node [style=none] (1) at (0, -0.5) {};
		\node [style=none] (2) at (0, 0.5) {};
	\end{pgfonlayer}
	\begin{pgfonlayer}{edgelayer}
		\draw [qWire] (2.center) to (0);
		\draw [qWire] (0) to (1.center);
	\end{pgfonlayer}
\end{tikzpicture}
$ . This is in this way that the process theory contains every possible decohered system from $\mathcal{P}$. Note, in particular, that this means that it contains $\mathcal{P}$ itself which is simply corresponds to restricting to systems of the form \resizebox{!}{3mm}{$\left(A,%
\InputIfFileExists{Diagrams/Karoubi4.tikz}{}{\input{./figures/Diagrams/Karoubi4.tikz}}\right)$}.

Typically it will be the case that $\mathcal{P}\subsetneq\mathcal{K[P]}$ and so we cannot directly use the Karoubi envelope to define a partitioning of $\mathcal{P}$. It is however relatively simple to see how one can modify $\mathcal{P}$ such that we can view $\mathcal{K[P]}$ as a set of decohered processes and leverage this to define a free partition and so on to resource theories. The way we do this is we extend $\mathcal{P}$ to give systems an additional label, namely, a decoherence process.

\begin{definition}[Decoherence labeled theory, $\mathcal{D}$]
Systems in $\mathcal{D[P]}$ are exactly the same as those in $\mathcal{K[P]}$, that is, they are labeled by arbitrary pairs:
\beq
\InputIfFileExists{Diagrams/Karoubi2.tikz}{}{\input{./figures/Diagrams/Karoubi2.tikz}}\resizebox{!}{3mm}{$\left(A,%
\InputIfFileExists{Diagrams/Karoubi1.tikz}{}{\input{./figures/Diagrams/Karoubi1.tikz}}\right)$}
\eeq
where $A$ is a system in $\mathcal{P}$ and $%
\InputIfFileExists{Diagrams/smallDecohere.tikz}{}{\input{./figures/Diagrams/smallDecohere.tikz}}$ is an idempotent on $A$. Composition similarly is the same as in $\mathcal{K[P]}$ that is:
\beq\label{eq:DPComp}
\InputIfFileExists{Diagrams/Karoubi2.tikz}{}{\input{./figures/Diagrams/Karoubi2.tikz}}\resizebox{!}{3mm}{$\left(A,%
\InputIfFileExists{Diagrams/Karoubi1.tikz}{}{\input{./figures/Diagrams/Karoubi1.tikz}}\right)$}
\ \
\InputIfFileExists{Diagrams/Karoubi2.tikz}{}{\input{./figures/Diagrams/Karoubi2.tikz}}\resizebox{!}{3mm}{$\left(B,%
\InputIfFileExists{Diagrams/Karoubi3.tikz}{}{\input{./figures/Diagrams/Karoubi3.tikz}}\right)$}
\quad := \quad
\InputIfFileExists{Diagrams/Karoubi2.tikz}{}{\input{./figures/Diagrams/Karoubi2.tikz}}\resizebox{!}{3mm}{$\left(AB,%
\InputIfFileExists{Diagrams/Karoubi1.tikz}{}{\input{./figures/Diagrams/Karoubi1.tikz}}%
\InputIfFileExists{Diagrams/Karoubi3.tikz}{}{\input{./figures/Diagrams/Karoubi3.tikz}}\right)$}
\eeq
The difference between $\mathcal{K[P]}$ and $\mathcal{D[P]}$ comes when we look at the processes. Specifically, processes in $\mathcal{D[P]}$ from  $\decA$ to $\decB$ are the full set of processes in $\mathcal{P}$ from $A$ to $B$ -- the decoherence map is simply a label and does not have an impact on the set of possible processes. These then compose exactly as their associated processes would in $\mathcal{P}$.
\end{definition}

That is, we take the same systems as in the Karoubi envelope but we do not restrict the processes to being the decohered processes, we allow them to be arbitrary processes from $\mathcal{P}$. It then immediately follows that $\mathcal{K[P]}\subseteq\mathcal{D[P]}$ and hence that this defines a partitioning of $\mathcal{D[P]}$. This, for the same reasons as discussed in Sec.~\ref{Sec:Decoherence}, will not lead to interesting resource theories of states and so we define a broader class of processes for the free subtheory.

Following our earlier discussions it is natural to define the analogue of $\mathsf{cDIO}$, however, note that the assumption about how systems compose in $\mathcal{D[P]}$ (as described by Eq.~\eqref{eq:DPComp}) and similarly in $\mathcal{K[P]}$ (as described by Eq.~\eqref{eq:KPComp}) means that we are in the situation described by Ex.~\ref{ex:LocalDecoherenceComposition} and so, as discussed just after Lem.~\ref{lem:34} it will be the case here that $\mathsf{cDIO}=\mathsf{DIO}$. We will therefore define the following as a potential set of free processes:
\begin{definition}
Given a decoherence labeled process theory $\mathcal{D[P]}$ we can define the subtheory $\mathsf{DIO}(=\mathsf{cDIO})$ as:
\beq
\mathsf{DIO} := \left\{\begin{tikzpicture}
	\begin{pgfonlayer}{nodelayer}
		\node [style=none] (0) at (0, 1.25) {};
		\node [style={small box}] (1) at (0, -0) {$g$};
		\node [style={right label}] (2) at (0, 0.75) {$\decB$};
		\node [style=none] (3) at (0, -1.25) {};
		\node [style={right label}] (4) at (0, -1) {$\decA$};
	\end{pgfonlayer}
	\begin{pgfonlayer}{edgelayer}
		\draw [qWire] (1) to (0.center);
		\draw [qWire] (3.center) to (1);
	\end{pgfonlayer}
\end{tikzpicture}
 \in \mathcal{D[P]}\ \middle|\ \begin{tikzpicture}
	\begin{pgfonlayer}{nodelayer}
		\node [style={small box}] (0) at (-0.5, -0) {$g$};
		\node [style=none] (1) at (-0.5000001, 2.25) {};
		\node [style=none] (2) at (-0.5, -1.25) {};
		\node [style={right label}] (3) at (-0.5, -1) {$\decA$};
		\node [style={right label}] (4) at (-0.5000001, 2) {$\decB$};
		\node [style={proj}] (5) at (-0.5000002, -1.5) {};
		\node [style=none] (6) at (-0.5, -1.75) {};
		\node [style={right label}] (7) at (-0.5, -2.25) {$\decA$};
		\node [style=none] (8) at (-0.5000001, -2.25) {};
	\end{pgfonlayer}
	\begin{pgfonlayer}{edgelayer}
		\draw [qWire] (2.center) to (0);
		\draw [qWire] (0) to (1.center);
		\draw [qWire] (8.center) to (6.center);
	\end{pgfonlayer}
\end{tikzpicture} =  \ \ \begin{tikzpicture}
	\begin{pgfonlayer}{nodelayer}
		\node [style={small box}] (0) at (-0.5, -0) {$g$};
		\node [style=none] (1) at (-0.5, 1.25) {};
		\node [style=none] (2) at (-0.5000001, -2.25) {};
		\node [style={right label}] (3) at (-0.5000001, -2.25) {$\decA$};
		\node [style={right label}] (4) at (-0.5, 0.75) {$\decB$};
		\node [style={proj},fill=black] (5) at (-0.5000002, 1.5) {};
		\node [style=none] (6) at (-0.5, 1.75) {};
		\node [style={right label}] (7) at (-0.5, 2) {$\decB$};
		\node [style=none] (8) at (-0.5, 2.25) {};
	\end{pgfonlayer}
	\begin{pgfonlayer}{edgelayer}
		\draw [qWire] (2.center) to (0);
		\draw [qWire] (0) to (1.center);
		\draw [qWire] (8.center) to (6.center);
	\end{pgfonlayer}
\end{tikzpicture}\ \ \right\}\eeq
\end{definition}

This defines a partitioned process theory $(\mathsf{DIO},\mathcal{D[P]})$, which can then be used to construct various resource theories of coherence following the methods outlined in App.~\ref{App:ResourceTheories}.

The resource theories which arise from this partitioned process theory will be much more elaborate than those considered so far in the literature. This is because they do not pick a particular decoherence mechanism for each system, but allow one to consider every possible decoherence mechanism for each system. Even in the simplest case of constructing a resource theory of states for this partitioned process theory, we find a substantially richer structure than has so far been considered in the literature -- developing a deeper understanding of this structure will be left for future work.

Note that it is possible to go beyond the assumption that global decoherence is simply the product of local decoherences which led to the fact that $\mathsf{cDIO}=\mathsf{DIO}$. To do so would involve redefining how systems compose within $\mathcal{K[P]}$ and $\mathcal{D[P]}$. That is, we would take some new composition rule:
\beq
\InputIfFileExists{Diagrams/Karoubi2.tikz}{}{\input{./figures/Diagrams/Karoubi2.tikz}}\resizebox{!}{3mm}{$\left(A,%
\InputIfFileExists{Diagrams/Karoubi1.tikz}{}{\input{./figures/Diagrams/Karoubi1.tikz}}\right)$}
\ \
\InputIfFileExists{Diagrams/Karoubi2.tikz}{}{\input{./figures/Diagrams/Karoubi2.tikz}}\resizebox{!}{3mm}{$\left(B,%
\InputIfFileExists{Diagrams/Karoubi3.tikz}{}{\input{./figures/Diagrams/Karoubi3.tikz}}\right)$}
\quad := \quad
\InputIfFileExists{Diagrams/Karoubi2.tikz}{}{\input{./figures/Diagrams/Karoubi2.tikz}}\resizebox{!}{3mm}{$\left(AB,\begin{tikzpicture}
	\begin{pgfonlayer}{nodelayer}
		\node [style={right label}] (0) at (-0.5000002, -0.7499999) {$A$};
		\node [style={right label}] (1) at (0.5000002, -0.7499999) {$B$};
		\node [style=none] (2) at (0.5000002, 0.7499999) {};
		\node [style=none] (3) at (0.5000002, 0.2500001) {};
		\node [style=none] (4) at (-0.5000002, 0.2500001) {};
		\node [style={right label}] (5) at (0.5000002, 0.5) {$B$};
		\node [style={wide proj}, fill=gray] (6) at (0, -0) {};
		\node [style={right label}] (7) at (-0.5000002, 0.5) {$A$};
		\node [style=none] (8) at (-0.5000002, 0.7499999) {};
		\node [style=none] (9) at (0.5000002, -0.7499999) {};
		\node [style=none] (10) at (-0.5000002, -0.7499999) {};
		\node [style=none] (11) at (-0.5000002, -0.2500001) {};
		\node [style=none] (12) at (0.5000002, -0.2500001) {};
	\end{pgfonlayer}
	\begin{pgfonlayer}{edgelayer}
		\draw [qWire] (8.center) to (4.center);
		\draw [qWire] (2.center) to (3.center);
		\draw [qWire] (11.center) to (10.center);
		\draw [qWire] (12.center) to (9.center);
	\end{pgfonlayer}
\end{tikzpicture}
\right)$}
\eeq
where the global decoherence
\beq
\begin{tikzpicture}
	\begin{pgfonlayer}{nodelayer}
		\node [style={right label}] (0) at (-0.5000002, -0.7499999) {$A$};
		\node [style={right label}] (1) at (0.5000002, -0.7499999) {$B$};
		\node [style=none] (2) at (0.5000002, 0.7499999) {};
		\node [style=none] (3) at (0.5000002, 0.2500001) {};
		\node [style=none] (4) at (-0.5000002, 0.2500001) {};
		\node [style={right label}] (5) at (0.5000002, 0.5) {$B$};
		\node [style={wide proj}, fill=gray] (6) at (0, -0) {};
		\node [style={right label}] (7) at (-0.5000002, 0.5) {$A$};
		\node [style=none] (8) at (-0.5000002, 0.7499999) {};
		\node [style=none] (9) at (0.5000002, -0.7499999) {};
		\node [style=none] (10) at (-0.5000002, -0.7499999) {};
		\node [style=none] (11) at (-0.5000002, -0.2500001) {};
		\node [style=none] (12) at (0.5000002, -0.2500001) {};
	\end{pgfonlayer}
	\begin{pgfonlayer}{edgelayer}
		\draw [qWire] (8.center) to (4.center);
		\draw [qWire] (2.center) to (3.center);
		\draw [qWire] (11.center) to (10.center);
		\draw [qWire] (12.center) to (9.center);
	\end{pgfonlayer}
\end{tikzpicture}
\eeq
satisfies certain constraints, for example, those discussed in Ex.~\ref{ex:GlobalDecoherenceComposition}. Again, we leave detailed analysis of this for future work.

Similarly, we can go beyond the assumption of speakable coherence to the more general unspeakable case. We simply must replace decoherence processes in the above constructions with decoherence mechanism -- i.e., in the end we will have one system per decoherence mechanism and a commuting condition defining the free processes. Once again, detailed analysis of this will be left for future work.
}
\section{Discussion}\label{Sec:Discussion}

In the previous section we illustrated the use of the process theoretic approach for constructing resource theories of coherence. As we explained,  this should be seen as a toolbox for constructing these resource theories: if you have a particular situation in mind then the above techniques should inform how to go about constructing the relevant resource theory, with the examples we have given serving as a guide. In constructing a resource theory of coherence there are many decisions that must be made:
\bit[noitemsep,topsep=0pt,parsep=0pt,partopsep=0pt]
\item[-] What are {the} decoherence processes, and, what compositionality constraints do {they} need to satisfy?
\item[-] {Is the coherence speakable or unspeakable}?
\item[-] If the answer is unspeakable coherence, then what are {the} decoherence mechanisms? e.g. how much and what information should be encoded in the environment?
\item[-] {Finally, what are the fundamental resources to be considered,} are they states, measurements, transformations or something more general?
\eit
Once these decisions have been made, then all of the machinery that we've introduced in this work can be employed to construct the {appropriate} resource theory.

Perhaps the most important lessons to be learnt here, which apply regardless of the above decisions that have to made, are that:
\bit[noitemsep,topsep=0pt,parsep=0pt,partopsep=0pt]
\item[-] One should demand that the free processes preserve the full set of decohered processes rather than just states---this rules out $\mathsf{MIO}$ as a suitable choice for free processes.
\item[-] Considering composite systems has an influence even on single-system resource theories---showing that we need to move from considering $\mathsf{DIO}$ processes to a `complete' notion which we term $\mathsf{cDIO}$.
\eit
These tell us that the largest set of free processes we should consider are the $\mathsf{cDIO}$ processes we defined.

Whilst $\mathsf{cDIO}$ processes are singled out as the largest set of free processes, if we want to consider unspeakable rather than speakable coherence, it is well motivated to consider a smaller set of free processes, we show the essential difference here comes when one considers the decoherence mechanism rather than just the decoherence process. This is closely related to the translationally invariant operations, $\mathsf{TIO}$, which have widely been considered in the literature, again with a caveat that they should be invariant in a complete sense, we call such processes $\mathsf{cTIO}$ processes. We find that this is in really a limiting case of the general construction which allows for us to interpolate between $\mathsf{cDIO}$ and $\mathsf{cTIO}$ by varying how much information is encoded into the environment system.

We have shown how we can, in an extremely general way, talk about resource theories of coherence for process theories. Interestingly, this makes no reference to tomography, convexity, or any of the usual structure of generalised probabilistic theories, it would therefore be interesting to know if coherence plays a role in fields which are normally well beyond the scope of what we consider in quantum foundations. Moreover, this challenges the standard tools used to study coherence in quantum information theory, no longer are bases playing a prominent role, or any convex or information theoretic structures. This opens the question as to how far we can push this approach, what else can we learn from this perspective?

\subsection{Future work}

There are plenty of directions for future work stemming from this, we hope this sparks interest in these tools and they are explored more broadly by the community.

Firstly we can consider applications of this formalism:
\bit[noitemsep,topsep=0pt,parsep=0pt,partopsep=0pt]
 \item[-] On the one hand, there is the question of applying this formalism specifically to quantum theory -- to do a more in depth study of the various resource theories that can be constructed, to understand the properties that they have, how these relate to information theoretic tasks, and how to define meaningful monotones to capture the preorder of resources. In particular, {i) to explore decoherence maps beyond the traditionally studied completely dephasing maps, ii) to explore resource theories with multiple sources of decoherence, and iii) to use these tools to}, in a methodical way, go beyond considering coherence of states to coherence of channels and more general objects.
\item[-] On the other hand, we can consider applying this formalism to other physical theories, for example, what can we say about coherence in generalised probabilistic theories \cite{barrett2007information}, in Spekkens toy model \cite{spekkens2007evidence}, or in more exotic process theories such as those described in \cite{gogioso2017fantastic}.
\eit

Secondly, we can consider further development of the formalism itself:
\bit[noitemsep,topsep=0pt,parsep=0pt,partopsep=0pt]
\item[-] In one direction, it would be beneficial to explore the relationship between all of the different resource theories that we can construct using these tools -- are there some basic principles which allow us to organise this `zoo' of resource theories? Moreover, given that we have multiple different ways to define the partitioned process theory, then it would be worth exploring the basic mathematical structure of this as a play ground for defining resource theories with multiple different resources. This should have connections to the work of \cite{sparaciari2018multi,sparaciari2018first}. {Indeed, section \ref{sec:multidecoherence} can be viewed as preliminary work in this direction -- at least for the case of decoherence.}
\item[-] In another direction we can consider generalising our notion of decoherence. At times it may be useful to consider decoherence not as being a physical process, but simply as some restriction on what some agent can do. For example viewing the rebit \cite{hardy2012limited} as a `decohered' form of a qubit. Such a picture strongly connects to the work of \cite{chiribella2018agents} and could be modelled by defining decoherence directly on the level of the process theory as an idempotent semi-functor. This would avoid the necessity for decoherence to be a physically implementable process and so, for example, this would allow one to view the rebit  as being a decohered form of a qubit. {More generally, if we do not demand that decoherence is physical then we would recover the notion of a \emph{resource destroying map} \cite{liu2017resource} our condition that a process is free relative to the resource destroying map would be a complete version of the commutativity condition in \cite{liu2017resource}.}
\eit

Finally, we hope that this sparks interest in the process-theoretic approach to resource theories, and {moreover} highlights the importance of {considering composition in a much more general sense than is} typically considered in the literature. We have seen that such consideration lead to important restrictions on the free processes for resource theories of coherence, and they may have important consequences for many other resource theories.

\section*{Acknowledgments}
Many thanks to Alex Wilce whose suggestions led to the development of section~\ref{sec:multidecoherence}. Thanks to Tom\'a\v{s} Gonda for helpful suggestions regarding generalising Example~\ref{ex:LocalDecoherenceComposition}. Thanks also to Rob Spekkens and Ana Bel\'en Sainz for helpful discussions. All diagrams were prepared using TikZit.

CML acknowledges funding from the EPSRC through the UCL EPSRC Doctoral Prize Fellowship. This research was supported in part by Perimeter Institute for Theoretical Physics. Research at Perimeter Institute is supported by the Government of Canada through the Department of Innovation, Science and Economic Development Canada and by the Province of Ontario through the Ministry of Research, Innovation and Science. This research was also supported in part by the Foundation for Polish Science through IRAP project co-financed by EU within
Smart Growth Operational Programme (contract no. 2018/MAB/5).

\bibliographystyle{plainurl}
\bibliography{bibliography}

\appendix

\section{Resource theories}\label{App:ResourceTheories}

{As discussed in the introduction, in Ref.~\cite{coecke2016mathematical}} a resource theory, $\mathcal{R}$, is defined to be formally equivalent to a process theory; however, rather than interpreting systems as being physical systems and processes as their evolution, we instead interpret systems as being resources, and processes as ways in which we can freely transform between resources. Composite systems simply represent having multiple different resources available. Sequential composition of the processes reflects the fact that if you can freely transform one resource into another and then freely transform that into a third, then there is a way to freely transform the first resource into the third resource. A similar story can be told for parallel composition. {A resource is then said to be free if there is some way to freely create the resource out of nothing.}

 To try to give some intuition for this distinction between process theories and resource theories, let us consider two different presentations of an LOCC transformation in the resource theory of entanglement. Specifically, let us consider the transformation of a maximally entangled Bell state into a product state by using local discard and prepare channels. On the one hand we can represent this in the process theory of quantum theory:
\beq%
\InputIfFileExists{Diagrams/LOCCProcessTheory.tikz}{}{\input{./figures/Diagrams/LOCCProcessTheory.tikz}}\eeq
where this diagram describes the physical procedure by which the Bell state is prepared, which is then composed with a discard and prepare $\ket{00}$ channel.

On the other hand we could consider the resource theory of LOCC-entanglement where the same procedure would be represented as:
\beq%
\InputIfFileExists{Diagrams/LOCCResourceTheory.tikz}{}{\input{./figures/Diagrams/LOCCResourceTheory.tikz}}\eeq
where this diagram should be read as saying that, if one had access to the resource of `a Bell state', then there is a free way to transform it under LOCC to the resource `a $\ket{00}$ state'.

{
There is, of course, a much simpler way to obtain the resource `a $\ket{00}$ state' from LOCC operations which is simply to directly prepare it. In our process theory of quantum theory this would be represented by:
\beq
\begin{tikzpicture}
	\begin{pgfonlayer}{nodelayer}
		\node [style=point] (0) at (-1, -0.5) {$0$};
		\node [style=point] (1) at (1, -0.5) {$0$};
		\node [style=none] (2) at (-1, 1) {};
		\node [style=none] (3) at (1, 1) {};
		\node [style={right label}] (4) at (-1, 0.5) {$\mathds{C}^2$};
		\node [style={right label}] (5) at (1, 0.5) {$\mathds{C}^2$};
	\end{pgfonlayer}
	\begin{pgfonlayer}{edgelayer}
		\draw [qWire] (2.center) to (0);
		\draw [qWire] (3.center) to (1);
	\end{pgfonlayer}
\end{tikzpicture}
\eeq
whilst in the resource theory of LOCC-entanglement this would be represented by:
\beq
\begin{tikzpicture}
	\begin{pgfonlayer}{nodelayer}
		\node [style={small box}] (0) at (0, -1) {Prepare $\ket{00}$};
		\node [style={right label}] (1) at (0, 0.5) {$\ket{00}$};
		\node [style=none] (2) at (0, 1) {};
	\end{pgfonlayer}
	\begin{pgfonlayer}{edgelayer}
		\draw (2.center) to (0);
	\end{pgfonlayer}
\end{tikzpicture}
\eeq
the fact that this resource can be constructed in the resource theory from `nothing' means that `a $\ket{00}$ state' is a \emph{free resource}.
}

In other words, whilst resource theories and process theories may be formally equivalent mathematical structures we should be extremely careful not to conflate them. There is however a strong connection that can be identified between these two perspectives, indeed the main contribution of \cite{coecke2016mathematical} was identifying the relevant structure that needs to be added to a process theory in order to construct a resource theory from it.

Specifically, given a process theory $\mathcal{P}$, in order to construct resource theories we must identify a sub-process theory of free processes $\mathcal{P}^{\mathsf{free}}$. The idea being that these are the physical processes which we can freely implement with no cost. For example, in constructing a resource theory of entanglement we may identify the LOCC operations as being free. This subset of free processes should be closed under composition, that is, if we can freely implement $f$ and freely implement $g$ then we should be able to freely do a composite of $f$ and $g$. Hence, rather than being an arbitrary subset of the processes this subset must be a sub process theory. Such a pair $(\mathcal{P},\mathcal{P}^{\mathsf{free}})$ is known as a \emph{partitioned process theory}.

Given a partitioned process theory $(\mathcal{P},\mathcal{P}^{\mathsf{free}})$ they show in \cite{coecke2016mathematical} that there are various different resource theories which can be constructed. Roughly speaking, this consists of identifying the sorts of processes within $\mathcal{P}$ which you want to consider to be resources. For example, do we want to think of just the states in $\mathcal{P}$ as being resources, or do we want to think of the channels as being resources? There are various constructions of this sort which are discussed in \cite{coecke2016mathematical}, we, for simplicity, will just discuss {a few} of these here. But it should be noted that this is not exhaustive, and neither was the original work. The precise construction that should be used will depend on exactly what one chooses to designate as a resource. In this sense, the partitioned process theory is really the more fundamental structure, and the particular resource theory that one constructs is situation dependent.

\begin{example}[Resource theory of states]\label{ex:RTofStates}Firstly let us construct a resource theory of \emph{states}. These are the most commonly considered resource theories within the literature. In such resource theories, the systems in the resource theory $\mathcal{R}$ correspond to all of the states in $\mathcal{P}$:
\beq
\left\{%
\InputIfFileExists{Diagrams/RTofStates.tikz}{}{\input{./figures/Diagrams/RTofStates.tikz}}\right\}\ \cong\ \left\{%
\InputIfFileExists{Diagrams/RTofStates2.tikz}{}{\input{./figures/Diagrams/RTofStates2.tikz}} \in \mathcal{P}
 \right\}
\eeq
where parallel composition of systems in $\mathcal{R}$ is given by parallel composition of the associated states in $\mathcal{P}$, that is:
\beq
\InputIfFileExists{Diagrams/RTS-ParallelSystem.tikz}{}{\input{./figures/Diagrams/RTS-ParallelSystem.tikz}} \ \sim \ %
\InputIfFileExists{Diagrams/RTS-ParallelSystem2.tikz}{}{\input{./figures/Diagrams/RTS-ParallelSystem2.tikz}}
\eeq
whilst the transformations correspond to those in $\mathcal{P}^{\mathsf{free}}$:
\beq
\left\{%
\InputIfFileExists{Diagrams/RTofStates3.tikz}{}{\input{./figures/Diagrams/RTofStates3.tikz}} \right\}\ \cong\ \left\{%
\InputIfFileExists{Diagrams/RTofStates4.tikz}{}{\input{./figures/Diagrams/RTofStates4.tikz}} \in \mathcal{P}^{\mathsf{free}}\ \middle|\ %
\InputIfFileExists{Diagrams/RTofStates5.tikz}{}{\input{./figures/Diagrams/RTofStates5.tikz}}\ = \ %
\InputIfFileExists{Diagrams/RTofStates6.tikz}{}{\input{./figures/Diagrams/RTofStates6.tikz}} \right\}
\eeq
that is, we can freely transform $s^A$ into $r^B$ if there is some free process with input $A$ and output $B$ that, in particular, maps the state $s$ of $A$ to the state $r$ of $B$.

 Composition of processes in $\mathcal{R}$ is given by the composition of processes in $\mathcal{P}$, in particular, sequential composition is given by:
\beq
\InputIfFileExists{Diagrams/RTS-SequentialProcess1.tikz}{}{\input{./figures/Diagrams/RTS-SequentialProcess1.tikz}} \ \sim \ \ %
\InputIfFileExists{Diagrams/RTS-SequentialProcess2.tikz}{}{\input{./figures/Diagrams/RTS-SequentialProcess2.tikz}}
\eeq
where $f$ transforms $s$ into $r$ and then $g$ transforms $r$ into $t$ in the process theory $\mathcal{P}$. Parallel composition is given by:
\beq
\InputIfFileExists{Diagrams/RTS-ParallelProcess1.tikz}{}{\input{./figures/Diagrams/RTS-ParallelProcess1.tikz}} \ \sim \ \ %
\InputIfFileExists{Diagrams/RTS-ParallelProcess2.tikz}{}{\input{./figures/Diagrams/RTS-ParallelProcess2.tikz}}
\eeq
where $f$ transforms $s$ into $r$ and $g$ transforms $t$ into $u$ within the process theory $\mathcal{P}$.

{The free resources are then simply the resources which can be constructed from nothing:
\beq
\left\{\begin{tikzpicture}
	\begin{pgfonlayer}{nodelayer}
		\node [style=none] (0) at (0, -1) {};
		\node [style=none] (1) at (0, 1) {};
		\node [style={right label}] (2) at (0, -0) {$s^A$};
	\end{pgfonlayer}
	\begin{pgfonlayer}{edgelayer}
		\draw (0.center) to (1.center);
	\end{pgfonlayer}
\end{tikzpicture}
\ \middle| \
 \begin{tikzpicture}
	\begin{pgfonlayer}{nodelayer}
		\node [style={small box}] (0) at (0, -0.5) {$s$};
		\node [style=none] (1) at (0, .75) {};
		\node [style={right label}] (2) at (0, .5) {$s^A$};
	\end{pgfonlayer}
	\begin{pgfonlayer}{edgelayer}
		\draw (0) to (1.center);
	\end{pgfonlayer}
\end{tikzpicture} \in \mathcal{R}
 \right\}
  \eeq
  which is clearly nothing but the set of states in the free partition as:
  \beq
   \left\{
 \begin{tikzpicture}
	\begin{pgfonlayer}{nodelayer}
		\node [style={small box}] (0) at (0, -0.5) {$s$};
		\node [style=none] (1) at (0, .75) {};
		\node [style={right label}] (2) at (0, .5) {$s^A$};
	\end{pgfonlayer}
	\begin{pgfonlayer}{edgelayer}
		\draw (0) to (1.center);
	\end{pgfonlayer}
\end{tikzpicture}
 \right\}
  \cong
   \left\{ \begin{tikzpicture}
	\begin{pgfonlayer}{nodelayer}
		\node [style=point] (0) at (0, -0.5) {$s$};
		\node [style=none] (1) at (0, .75) {};
		\node [style={right label}] (2) at (0, .5) {$A$};
	\end{pgfonlayer}
	\begin{pgfonlayer}{edgelayer}
		\draw [qWire] (0) to (1.center);
	\end{pgfonlayer}
\end{tikzpicture} \in \mathcal{P}^{\mathsf{free}}
 \right\}
 \eeq
}
\end{example}
{
\begin{example}[Resource theory of effects]
Recently there has been interested in resource theories of effects \cite{oszmaniec2019operational,guff2019resource}. We can capture that in our formalism as follows. Define the resources to simply be the effects in $\mathcal{P}$:
\beq
\left\{\begin{tikzpicture}
	\begin{pgfonlayer}{nodelayer}
		\node [style=none] (0) at (0, -1) {};
		\node [style=none] (1) at (0, 1) {};
		\node [style={right label}] (2) at (0, -0) {$e_A$};
	\end{pgfonlayer}
	\begin{pgfonlayer}{edgelayer}
		\draw (0.center) to (1.center);
	\end{pgfonlayer}
\end{tikzpicture}
\right\}\ \cong\ \left\{\begin{tikzpicture}
	\begin{pgfonlayer}{nodelayer}
		\node [style=copoint] (0) at (0, 0.5) {$e$};
		\node [style=none] (1) at (0, 0.25) {};
		\node [style=none] (2) at (0, -0.75) {};
		\node [style={right label}] (3) at (0, -0.5) {$A$};
	\end{pgfonlayer}
	\begin{pgfonlayer}{edgelayer}
		\draw [qWire] (2.center) to (1.center);
	\end{pgfonlayer}
\end{tikzpicture}
 \in \mathcal{P}
 \right\}
\eeq
where parallel composition of systems in $\mathcal{R}$ is given by parallel composition of the associated effects in $\mathcal{P}$, that is:
\beq \begin{tikzpicture}
	\begin{pgfonlayer}{nodelayer}
		\node [style=none] (0) at (-0.5, -1) {};
		\node [style=none] (1) at (-0.5, 1) {};
		\node [style={right label}] (2) at (-0.5, -0) {$e_A$};
		\node [style=none] (3) at (1, -1) {};
		\node [style=none] (4) at (1, 1) {};
		\node [style={right label}] (5) at (1, -0) {$e'_B$};
	\end{pgfonlayer}
	\begin{pgfonlayer}{edgelayer}
		\draw (0.center) to (1.center);
		\draw (3.center) to (4.center);
	\end{pgfonlayer}
\end{tikzpicture}
\sim\ \  \begin{tikzpicture}
	\begin{pgfonlayer}{nodelayer}
		\node [style=copoint] (0) at (-0.75, 0.5) {$e$};
		\node [style=none] (1) at (-0.75, 0.25) {};
		\node [style=none] (2) at (-0.75, -0.75) {};
		\node [style={right label}] (3) at (-0.75, -0.5) {$A$};
		\node [style=none] (4) at (0.75, -0.75) {};
		\node [style=none] (5) at (0.75, 0.25) {};
		\node [style=copoint] (6) at (0.75, 0.5) {$e'$};
		\node [style={right label}] (7) at (0.75, -0.5) {$B$};
	\end{pgfonlayer}
	\begin{pgfonlayer}{edgelayer}
		\draw [qWire] (2.center) to (1.center);
		\draw [qWire] (4.center) to (5.center);
	\end{pgfonlayer}
\end{tikzpicture}\eeq
transformations in $\mathcal{R}$ then correspond to processes in $\mathcal{P}^{\mathsf{free}}$ transforming effects into effects by pre-composition:
\beq \left\{\begin{tikzpicture}
	\begin{pgfonlayer}{nodelayer}
		\node [style={small box}] (0) at (0, -0) {$f$};
		\node [style=none] (1) at (0, 1.25) {};
		\node [style={right label}] (2) at (0, 1) {$e_A$};
		\node [style=none] (3) at (0, -1.25) {};
		\node [style=right label] (4) at (0, -1) {$e'_B$};
	\end{pgfonlayer}
	\begin{pgfonlayer}{edgelayer}
		\draw (0) to (1.center);
		\draw (3.center) to (0);
	\end{pgfonlayer}
\end{tikzpicture}
\right\}
\ \ \cong \ \
 \left\{
\InputIfFileExists{Diagrams/RTofStates4.tikz}{}{\input{./figures/Diagrams/RTofStates4.tikz}} \in \mathcal{P}^{\mathsf{free}}\ \middle|\ \begin{tikzpicture}
	\begin{pgfonlayer}{nodelayer}
		\node [style=copoint] (0) at (0, 1.5) {$e'$};
		\node [style=none] (1) at (0, 1.25) {};
		\node [style={right label}] (2) at (0, 0.75) {$B$};
		\node [style=none] (3) at (0.5, 0.25) {};
		\node [style=none] (4) at (0, -0.75) {};
		\node [style=none] (5) at (-0.5, -0.75) {};
		\node [style=none] (6) at (0, 0.25) {};
		\node [style=none] (7) at (0, -0.25) {$f$};
		\node [style=none] (8) at (-0.5, 0.25) {};
		\node [style=none] (9) at (0.5, -0.75) {};
		\node [style=none] (10) at (0, -1.5) {};
		\node [style={right label}] (11) at (0, -1.25) {$A$};
	\end{pgfonlayer}
	\begin{pgfonlayer}{edgelayer}
		\draw (8.center) to (3.center);
		\draw (3.center) to (9.center);
		\draw (9.center) to (5.center);
		\draw (5.center) to (8.center);
		\draw [qWire] (1.center) to (6.center);
		\draw [qWire] (4.center) to (10.center);
	\end{pgfonlayer}
\end{tikzpicture}
\ = \
\begin{tikzpicture}
	\begin{pgfonlayer}{nodelayer}
		\node [style=copoint] (0) at (0, 0.5) {$e$};
		\node [style=none] (1) at (0, 0.25) {};
		\node [style=none] (2) at (0, -0.75) {};
		\node [style={right label}] (3) at (0, -0.5) {$A$};
	\end{pgfonlayer}
	\begin{pgfonlayer}{edgelayer}
		\draw [qWire] (2.center) to (1.center);
	\end{pgfonlayer}
\end{tikzpicture}
\right\}
\eeq
Composition of processes in $\mathcal{R}$ is given by the relevant composition of processes in $\mathcal{P}$.

{Similarly to the case of states, it is easy to show that the free resources in $\mathcal{R}$ will be in one-to-one correspondence with the effects in $\mathcal{P}^{\mathsf{free}}$.
}

\end{example}

We can now develop a more complex example here, in which we consider measurements rather than just effects, in which case they can be freely transformed by suitable pre and post processing.
}
\begin{example}[Resource theory of measurements]\label{ex:RTofMeas}
Recently there has been interested in resource theories of measurements \cite{oszmaniec2019operational,guff2019resource}. If our process theory has a suitable notion of measurement, that is, if it contains classical systems, then we could define this in its own right. Define the resources in $\mathcal{R}$ to correspond to all of the measurements in $\mathcal{P}$:
\beq
\left\{%
\InputIfFileExists{Diagrams/RTofMeasurements.tikz}{}{\input{./figures/Diagrams/RTofMeasurements.tikz}}\right\}\ \cong\ \left\{%
\InputIfFileExists{Diagrams/destructiveMeasurement.tikz}{}{\input{./figures/Diagrams/destructiveMeasurement.tikz}} \in \mathcal{P}
 \right\}
\eeq
where parallel composition of systems in $\mathcal{R}$ is given by parallel composition of the associated measurements in $\mathcal{P}$, that is:
\beq %
\InputIfFileExists{Diagrams/RTofMeasurementsParallel.tikz}{}{\input{./figures/Diagrams/RTofMeasurementsParallel.tikz}} \sim\ \  %
\InputIfFileExists{Diagrams/destructiveMeasurementParallel.tikz}{}{\input{./figures/Diagrams/destructiveMeasurementParallel.tikz}}\eeq
transformations in $\mathcal{R}$ then correspond to pairs of processes in $\mathcal{P}^{\mathsf{free}}$ transforming measurements into measurements by pre and post composition:
\beq \left\{%
\InputIfFileExists{Diagrams/RTofMeasurements3.tikz}{}{\input{./figures/Diagrams/RTofMeasurements3.tikz}}\right\} \ \ \cong \ \  \left\{\begin{tikzpicture}
	\begin{pgfonlayer}{nodelayer}
		\node [style=none] (0) at (0, 1.25) {};
		\node [style=none] (1) at (0, 2.25) {};
		\node [style=none] (2) at (2, 2.25) {};
		\node [style=none] (3) at (2, 1.25) {};
		\node [style=none] (4) at (2, -2.25) {};
		\node [style=none] (5) at (2, -1.25) {};
		\node [style=none] (6) at (0, -1.25) {};
		\node [style=none] (7) at (0, -2.25) {};
		\node [style=none] (8) at (1.5, 1.25) {};
		\node [style=none] (9) at (1.5, -1.25) {};
		\node [style=none] (10) at (0.5, -1.25) {};
		\node [style=none] (11) at (0.5, -0.5) {};
		\node [style=none] (12) at (1, -2.25) {};
		\node [style=none] (13) at (1, -3) {};
		\node [style=none] (14) at (1, -1.75) {$f_{\mathsf{pre}}$};
		\node [style=none] (15) at (1, 1.75) {$f_{\mathsf{post}}$};
		\node [style={right label}] (16) at (1.5, -0) {$E$};
		\node [style={right label}] (17) at (1, -3) {$A$};
		\node [style={right label}] (18) at (0.5, -0.5) {$B$};
		\node [style={right label}] (19) at (0.5, 0.5) {$n$};
		\node [style={right label}] (20) at (1, 2.75) {$m$};
		\node [style=none] (21) at (0.5, 1.25) {};
		\node [style=none] (22) at (0.5, 0.5) {};
		\node [style=none] (23) at (1, 3) {};
		\node [style=none] (24) at (1, 2.25) {};
	\end{pgfonlayer}
	\begin{pgfonlayer}{edgelayer}
		\draw (1.center) to (2.center);
		\draw (2.center) to (3.center);
		\draw (3.center) to (0.center);
		\draw (0.center) to (1.center);
		\draw (6.center) to (5.center);
		\draw (5.center) to (4.center);
		\draw (4.center) to (7.center);
		\draw (7.center) to (6.center);
		\draw [qWire] (13.center) to (12.center);
		\draw [qWire] (10.center) to (11.center);
		\draw [qWire] (9.center) to (8.center);
		\draw [cWire] (21.center) to (22.center);
		\draw [cWire] (23.center) to (24.center);
	\end{pgfonlayer}
\end{tikzpicture} \in \mathcal{P}^{\mathsf{free}}\ \middle|\ %
\InputIfFileExists{Diagrams/destructiveMeasurementTransformed.tikz}{}{\input{./figures/Diagrams/destructiveMeasurementTransformed.tikz}}\ = \ %
\InputIfFileExists{Diagrams/destructiveMeasurement.tikz}{}{\input{./figures/Diagrams/destructiveMeasurement.tikz}} \right\}\eeq
Composition of processes in $\mathcal{R}$ is given by the relevant composition of processes in $\mathcal{P}$, as will be explained in more detail in Ex.~\ref{ex:RTofProc}.

{
Recalling that a resources is free if it can be constructed from nothing, it is clear that they are in one-to-one correspondence with diagrams of the form:
\beq
\begin{tikzpicture}
	\begin{pgfonlayer}{nodelayer}
		\node [style=none] (0) at (0, 1.25) {};
		\node [style=none] (1) at (0, 2.25) {};
		\node [style=none] (2) at (2, 2.25) {};
		\node [style=none] (3) at (2, 1.25) {};
		\node [style=none] (4) at (2, -2.25) {};
		\node [style=none] (5) at (2, -1.25) {};
		\node [style=none] (6) at (0, -1.25) {};
		\node [style=none] (7) at (0, -2.25) {};
		\node [style=none] (8) at (1.5, 1.25) {};
		\node [style=none] (9) at (1.5, -1.25) {};
		\node [style=none] (10) at (0.5, -1.25) {};
		\node [style=none] (11) at (1, -2.25) {};
		\node [style=none] (12) at (1, -3) {};
		\node [style=none] (13) at (1, -1.75) {$f_{\mathsf{pre}}$};
		\node [style=none] (14) at (1, 1.75) {$f_{\mathsf{post}}$};
		\node [style={right label}] (15) at (1.5, -0) {$E$};
		\node [style={right label}] (16) at (1, -3) {$A$};
		\node [style={right label}] (17) at (1, 2.75) {$m$};
		\node [style=none] (18) at (0.5, 1.25) {};
		\node [style=none] (19) at (1, 3) {};
		\node [style=none] (20) at (1, 2.25) {};
	\end{pgfonlayer}
	\begin{pgfonlayer}{edgelayer}
		\draw (1.center) to (2.center);
		\draw (2.center) to (3.center);
		\draw (3.center) to (0.center);
		\draw (0.center) to (1.center);
		\draw (6.center) to (5.center);
		\draw (5.center) to (4.center);
		\draw (4.center) to (7.center);
		\draw (7.center) to (6.center);
		\draw [qWire] (12.center) to (11.center);
		\draw [qWire] (9.center) to (8.center);
		\draw [cWire] (19.center) to (20.center);
	\end{pgfonlayer}
\end{tikzpicture}
\eeq
where $f_{\mathsf{pre}}, f_{\mathsf{post}} \in \mathcal{P}^{\mathsf{free}}$. As $\mathcal{P}^{\mathsf{free}}$ is closed under composition it is clear that this is nothing but a measurement in $\mathcal{P}^{\mathsf{free}}$ and, moreover, if we take the case that $E=A$ and $f_{\mathsf{pre}} =\mathds{1}_A$ it is clear that any measurement in $\mathcal{P}^{\mathsf{free}}$ can be written in this form. Hence, the free resources are in one-to-one correspondence with the measurements in the free partition.

}
\end{example}
\begin{example}[Resource theory of processes]\label{ex:RTofProc}
{Finally, let us construct a resource theory of general processes. This is less commonly considered in the literature than resource theories of states, the ability for this general compositional framework to so naturally handle this situation is one of its main advantages, there are some notable exceptions however such as \cite{liu2020operational,liu2019resource,gour2019quantify}.} The first difference is that now we label the systems in the resource theory $\mathcal{R}$ by arbitrary processes in $\mathcal{P}$, that is:
\beq
\left\{%
\InputIfFileExists{Diagrams/RTofProcesses.tikz}{}{\input{./figures/Diagrams/RTofProcesses.tikz}}\right\}\ \cong\ \left\{%
\InputIfFileExists{Diagrams/RTofProcesses2.tikz}{}{\input{./figures/Diagrams/RTofProcesses2.tikz}} \in \mathcal{P}
 \right\}
\eeq
with parallel composition of systems being given by parallel composition of the associated processes. Transformations in the resource theory are then given by `free 1-combs', $\xi=(\xi_1\in\mathcal{P}^{\mathsf{free}},\xi_2\in\mathcal{P}^{\mathsf{free}})$, that is:
\beq
\left\{%
\InputIfFileExists{Diagrams/RTofProcesses3.tikz}{}{\input{./figures/Diagrams/RTofProcesses3.tikz}} \right\}\ \cong\ \left\{%
\InputIfFileExists{Diagrams/RTofProcesses4.tikz}{}{\input{./figures/Diagrams/RTofProcesses4.tikz}}\in\mathcal{P}^{\mathsf{free}} \middle|\ %
\InputIfFileExists{Diagrams/RTofProcesses5.tikz}{}{\input{./figures/Diagrams/RTofProcesses5.tikz}}\ = \ %
\InputIfFileExists{Diagrams/RTofProcesses6.tikz}{}{\input{./figures/Diagrams/RTofProcesses6.tikz}} \right\}
\eeq
We can then define sequential composition as:
\beq
\InputIfFileExists{Diagrams/RTP-SequentialProcess1.tikz}{}{\input{./figures/Diagrams/RTP-SequentialProcess1.tikz}}\ \sim\ %
\InputIfFileExists{Diagrams/RTofProcessesSequential.tikz}{}{\input{./figures/Diagrams/RTofProcessesSequential.tikz}}
\eeq
where $\xi$ maps $f$ to $g$ and then $\eta$ maps $g$ to $h$ within $\mathcal{P}$. Parallel composition is defined as:
\beq
\InputIfFileExists{Diagrams/RTP-ParallelProcess1.tikz}{}{\input{./figures/Diagrams/RTP-ParallelProcess1.tikz}}\ \sim\ %
\InputIfFileExists{Diagrams/RTofProcessesParallel.tikz}{}{\input{./figures/Diagrams/RTofProcessesParallel.tikz}}
\eeq
where $\xi$ maps $f$ to $g$ and $\eta$ maps $h$ to $i$ within $\mathcal{P}$.

{By the same logic as in Ex.~\ref{ex:RTofMeas} it can be shown that the free resources are in one-to-one correspondence with the set of processes in $\mathcal{P}^{\mathsf{free}}$.}
\end{example}

Whilst this last example may seem very general, there is actually a substantial limitation to this approach. Namely, that it specifies that all of the resources must be processed in parallel. {That is, there is no way to take two processes viewed as resources and to compose them in sequence to obtain a new resource.

\begin{example}[Resource theory of sets of processes]\label{ex:RTofProcSet}
A resource in this theory will be labeled by an indexed set of processes in $\mathcal{P}$:
\beq
\begin{tikzpicture}
	\begin{pgfonlayer}{nodelayer}
		\node [style=none] (0) at (0, -1) {};
		\node [style=none] (1) at (0, 1) {};
		\node [style={right label}] (2) at (0, -0) {$\{f_i \in \mathcal{P}\}_{i\in I}$};
	\end{pgfonlayer}
	\begin{pgfonlayer}{edgelayer}
		\draw (0.center) to (1.center);
	\end{pgfonlayer}
\end{tikzpicture}
\eeq
and parallel composition of these resources will be given by the disjoint union of the sets:
\beq
\{f_i\}_{i\in I} \otimes \{g_j\}_{j\in J}:= \{f_i\}_{i\in I} \sqcup \{g_j\}_{j\in J}
\eeq
For a full formalisation of this see \cite{coecke2016mathematical}, intuitively, however a free transformation $\xi: \{f_i\}_{i\in I}  \to \{g_j\}_{j\in J}$ is given by:
\bit
\item a disjoint partitioning of $I = \bigsqcup_{j\in J} I_j$
\item for each $j\in J$ a circuit $\xi_j$ constructed out of processes in $\mathcal{P}^{\mathsf{free}}$ such that
    \bit
    \item there is a `hole' in the circuit for each $f_i$ where $i\in I_j$, and
    \item when all of the $f_i$ are `plugged into these holes' the resulting process is $g_j$
    \eit
\eit
One can then show that the free resources in such a resource theory are simply given by:
\beq
\begin{tikzpicture}
	\begin{pgfonlayer}{nodelayer}
		\node [style=none] (0) at (0, -1) {};
		\node [style=none] (1) at (0, 1) {};
		\node [style={right label}] (2) at (0, -0) {$\{f_i \in \mathcal{P}^{\mathsf{free}}\}_{i\in I}$};
	\end{pgfonlayer}
	\begin{pgfonlayer}{edgelayer}
		\draw (0.center) to (1.center);
	\end{pgfonlayer}
\end{tikzpicture}
\eeq
\end{example}

So far we have seen that we define free processings in terms of higher-order processes, such as the combs in \ref{ex:RTofProc} and the circuits with holes in \ref{ex:RTofProcSet}, we can also consider resource theories in which the resources themselves are higher-order transformations, for example:

\begin{example}[Resource theory of $2$-combs]
Let us consider resources to be combs of the shape:
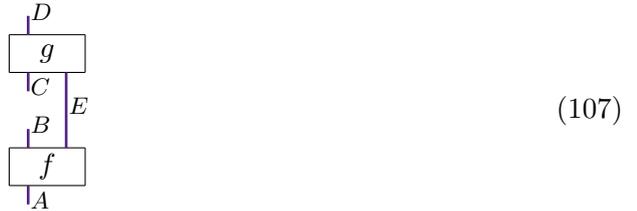
\beq%
\InputIfFileExists{Diagrams/comb.tikz}{}{\input{./figures/Diagrams/comb.tikz}}\eeq
that is, we will label the wires of the resource theory by these:
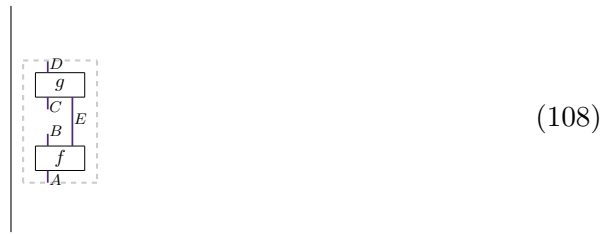
\beq
\begin{tikzpicture}
	\begin{pgfonlayer}{nodelayer}
		\node [style=none] (0) at (0, -3) {};
		\node [style=none] (1) at (0, 3) {};
		\node [style={right label}] (2) at (-1, 1) {};
	\end{pgfonlayer}
	\begin{pgfonlayer}{edgelayer}
		\draw (0.center) to (1.center);
	\end{pgfonlayer}
\end{tikzpicture}\
{\resizebox{10mm}{!}{$%
\InputIfFileExists{Diagrams/combLabel.tikz}{}{\input{./figures/Diagrams/combLabel.tikz}}$}}
\eeq
These have been far less widely studied in the literature, however \cite{ducuara2020multi} can be viewed as a special case of this resource theory in which $E$ is assumed to be trivial, $f$ is assumed to be a state and $g$ is assumed to be a measurement.

Parallel composition of resources is given by:
\beq
\begin{tikzpicture}
	\begin{pgfonlayer}{nodelayer}
		\node [style=none] (0) at (0, -3) {};
		\node [style=none] (1) at (0, 3) {};
		\node [style={right label}] (2) at (-1, 1) {};
	\end{pgfonlayer}
	\begin{pgfonlayer}{edgelayer}
		\draw (0.center) to (1.center);
	\end{pgfonlayer}
\end{tikzpicture}\
{\resizebox{10mm}{!}{$%
\InputIfFileExists{Diagrams/combLabel.tikz}{}{\input{./figures/Diagrams/combLabel.tikz}}$}}
\begin{tikzpicture}
	\begin{pgfonlayer}{nodelayer}
		\node [style=none] (0) at (0, -3) {};
		\node [style=none] (1) at (0, 3) {};
		\node [style={right label}] (2) at (-1, 1) {};
	\end{pgfonlayer}
	\begin{pgfonlayer}{edgelayer}
		\draw (0.center) to (1.center);
	\end{pgfonlayer}
\end{tikzpicture}\
{\resizebox{10mm}{!}{$%
\InputIfFileExists{Diagrams/combLabel2.tikz}{}{\input{./figures/Diagrams/combLabel2.tikz}}$}}
\ = \
\begin{tikzpicture}
	\begin{pgfonlayer}{nodelayer}
		\node [style=none] (0) at (0, -3) {};
		\node [style=none] (1) at (0, 3) {};
		\node [style={right label}] (2) at (-1, 1) {};
	\end{pgfonlayer}
	\begin{pgfonlayer}{edgelayer}
		\draw (0.center) to (1.center);
	\end{pgfonlayer}
\end{tikzpicture}\
{\resizebox{17mm}{!}{$%
\InputIfFileExists{Diagrams/combLabel3.tikz}{}{\input{./figures/Diagrams/combLabel3.tikz}}$}}
\eeq
The free transformations, $\xi$, within the resource theory are in one-to-one correspondence with particular diagrams constructed out of processes in $\mathcal{P}^{\mathsf{free}}$, for example, diagrams of the forms:
\beq
\begin{tikzpicture}
	\begin{pgfonlayer}{nodelayer}
		\node [style={small box}] (0) at (0, -0) {$\xi$};
		\node [style=none] (1) at (0, 1.25) {};
		\node [style=none] (3) at (0, -1.25) {};
	\end{pgfonlayer}
	\begin{pgfonlayer}{edgelayer}
		\draw (0) to (1.center);
		\draw (3.center) to (0);
	\end{pgfonlayer}
\end{tikzpicture}
\quad\sim\quad%
\InputIfFileExists{Diagrams/combTransformation.tikz}{}{\input{./figures/Diagrams/combTransformation.tikz}}\quad\text{,}\quad %
\InputIfFileExists{Diagrams/combTransformation2.tikz}{}{\input{./figures/Diagrams/combTransformation2.tikz}}\quad\text{or}\quad %
\InputIfFileExists{Diagrams/combTransformation4.tikz}{}{\input{./figures/Diagrams/combTransformation4.tikz}}
\eeq
where $\xi_i \in \mathcal{P}^{\mathsf{free}}$, correspond to transformations for the resource theory $\mathcal{R}$.

These,for example, can be composed in sequence as:
\beq
\InputIfFileExists{Diagrams/combTransformationSequence.tikz}{}{\input{./figures/Diagrams/combTransformationSequence.tikz}}
\eeq
and in parallel as:
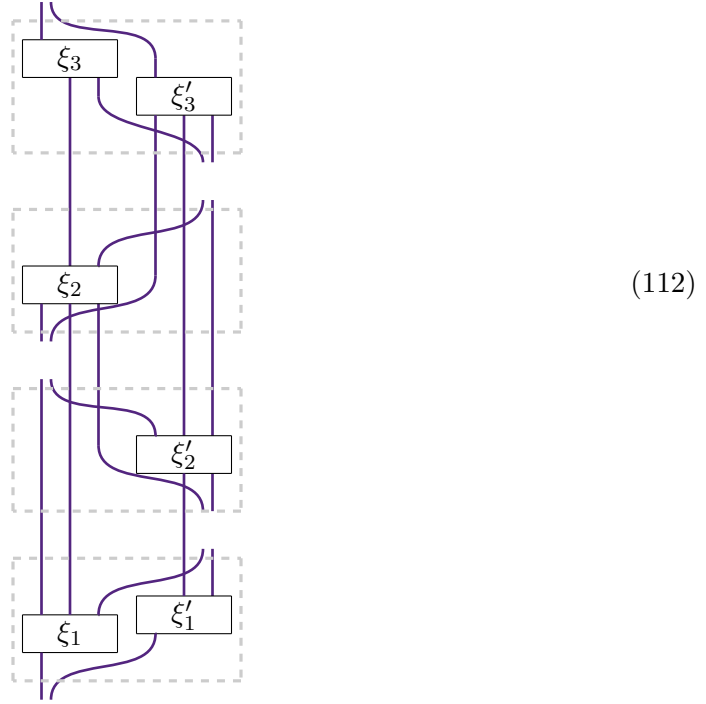
\beq
\InputIfFileExists{Diagrams/combTransformationParallel.tikz}{}{\input{./figures/Diagrams/combTransformationParallel.tikz}}
\eeq
Finally, the free resources are those that can be freely constructed from nothing, that is, resources:
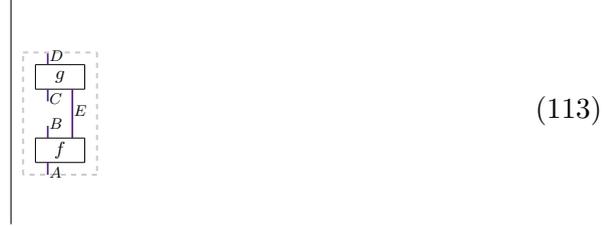
\beq
\begin{tikzpicture}
	\begin{pgfonlayer}{nodelayer}
		\node [style=none] (0) at (0, -3) {};
		\node [style=none] (1) at (0, 3) {};
		\node [style={right label}] (2) at (-1, 1) {};
	\end{pgfonlayer}
	\begin{pgfonlayer}{edgelayer}
		\draw (0.center) to (1.center);
	\end{pgfonlayer}
\end{tikzpicture}\
{\resizebox{10mm}{!}{$%
\InputIfFileExists{Diagrams/combLabel.tikz}{}{\input{./figures/Diagrams/combLabel.tikz}}$}}
\eeq
in which $f,g \in \mathcal{P}^{\mathsf{free}}$.
\end{example}
This is just a few examples of the sorts of resource theories that can be constructed from a partitioned process theory. It is clearly impossible to make an exhaustive list of examples as there is an infinite number of different sorts of things that one may want to consider as resources depending on the physical situation on hand. However, in future work it would be interesting to explore the scope of such resource theories and to develop a formalism which handles the construction of any resource theory from a partitioned process theory in a uniform and principled way.
}

{As mentioned in the introduction, in} the study of resource theories we are often not interested in the details of how we transform one resource into another, but simply whether or not there exists some way to freely transform one resource into another. That is, we often work simply with the preorder, $R$, that these resource theories, $\mathcal{R}$, define. For example, in the case of the resource theory of states, we work with the preorder defined by:
\beq
s^A \succ_R r^B \quad \iff \quad \exists\ \left(f:s^A \to r^B\right) \ \in \mathcal{R}
\eeq
{recall that this means that $f\in \mathcal{P}^{\mathsf{free}}$ within the partitioned process theory. Similarly,}
in the resource theory of processes we work with the preorder defined by:
\beq
f_A^B \succ_R g_C^D \quad \iff \quad \exists\ \left(\xi:f_A^B \to g_C^D\right)\ \in \mathcal{R}
\eeq
{recall also that this means that $\xi$ can be viewed a `comb' constructed out of processes in $\mathcal{P}^{\mathsf{free}}$. Essentially, in order to move from the resource theory, $\mathcal{R}$, to the preorder, $R$, we simply}
forget all of details of the resource theory except for whether the set of transformations from one resource to another is the empty set or not.

Often we are not able to efficiently characterise this preorder, in which case the study of resource theories often revolves around the construction of \emph{monotones}. That is, a function $\alpha:R \to \mathds{R}$ such that:
\beq
s \succ_R r \implies \alpha(s) \geq \alpha(r)
\eeq
These usual aim for such monotones is to find out some information about the preorder, in a way that is simple to compute and has some physical or operational significance. For this work we simply aim for an `in principle' characterisation of $\mathcal{R}$ and hence $R$ rather than trying to find efficient characterisations of this structure via monotones.

\section{Proof of lemma \ref{lem:23}}\label{app:23}
\proof
To see that this enforces that the composition of decohered is itself a decohered process it suffices to consider a pair of decohered processes $f$ and $g$ composed as:
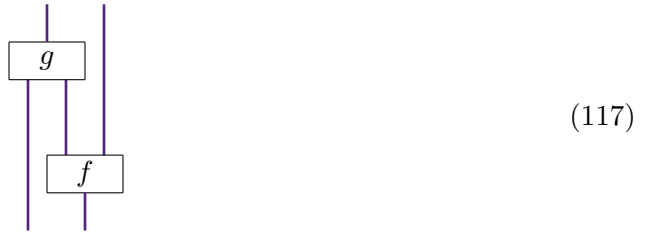
\beq%
\InputIfFileExists{Diagrams/DecComp1.tikz}{}{\input{./figures/Diagrams/DecComp1.tikz}}\eeq
as all other ways of composing processes can be seen as special cases of this. Then, as they are decohered we can write that:
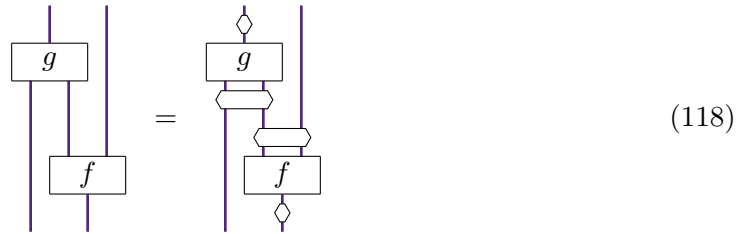
\beq%
\InputIfFileExists{Diagrams/DecComp1.tikz}{}{\input{./figures/Diagrams/DecComp1.tikz}}\ \ = \ \ %
\InputIfFileExists{Diagrams/DecComp2.tikz}{}{\input{./figures/Diagrams/DecComp2.tikz}}\eeq
Then using constraint \eqref{eq:LocalDecoherence} together with idempotence of decoherence processes we find that:
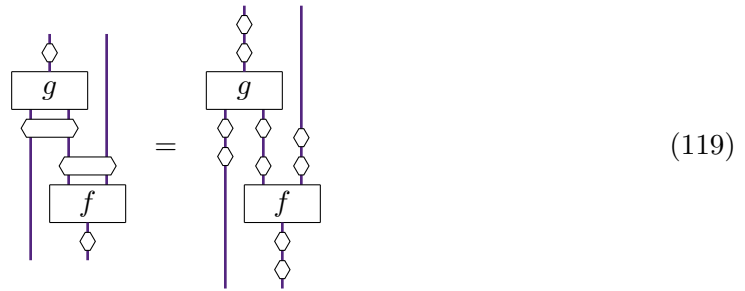
\beq%
\InputIfFileExists{Diagrams/DecComp2.tikz}{}{\input{./figures/Diagrams/DecComp2.tikz}}\ \ = \ \ %
\InputIfFileExists{Diagrams/DecComp3.tikz}{}{\input{./figures/Diagrams/DecComp3.tikz}}\eeq
Then, using the fact that $f$ and $g$ are decohered for a second time, and a simple diagrammatic rewriting, we obtain:
\beq%
\InputIfFileExists{Diagrams/DecComp3.tikz}{}{\input{./figures/Diagrams/DecComp3.tikz}}\ \ = \ \ %
\InputIfFileExists{Diagrams/DecComp4.tikz}{}{\input{./figures/Diagrams/DecComp4.tikz}}\eeq
Finally, using our constraint on composition of decoherences, eq.~\eqref{eq:LocalDecoherence},  we can write that:
\beq%
\InputIfFileExists{Diagrams/DecComp4.tikz}{}{\input{./figures/Diagrams/DecComp4.tikz}}\ \ = \ \ %
\InputIfFileExists{Diagrams/DecComp5.tikz}{}{\input{./figures/Diagrams/DecComp5.tikz}}\eeq
From which we can conclude that this composite of $f$ and $g$ is also a decohered process.
\endproof

\section{Proof of lemma \ref{lem:24}} \label{app:24}
\proof
Firstly consider the RHS of the first constraint, eq.~\eqref{eq:24}, and note that, by applying the second constraint (eq.~\eqref{eq:25}) multiple times, we can write that:
\beq%
\InputIfFileExists{Diagrams/DecComp6.tikz}{}{\input{./figures/Diagrams/DecComp6.tikz}}\ \ = \ \ %
\InputIfFileExists{Diagrams/DecComp7.tikz}{}{\input{./figures/Diagrams/DecComp7.tikz}} \ \ = \ \ %
\InputIfFileExists{Diagrams/DecComp8.tikz}{}{\input{./figures/Diagrams/DecComp8.tikz}}\eeq
Equating the left hand side using our first constraint (eq.~\eqref{eq:24}) and using idempotence of decoherence processes on the right hand side then gives us:
\beq \label{eq:C2}%
\InputIfFileExists{Diagrams/DecComp10.tikz}{}{\input{./figures/Diagrams/DecComp10.tikz}}\ \ = \ \ %
\InputIfFileExists{Diagrams/DecComp9.tikz}{}{\input{./figures/Diagrams/DecComp9.tikz}} \eeq
Now, consider the composite of two decohered processes $f$ and $g$, the fact that they are decohered means that we can write:
\beq%
\InputIfFileExists{Diagrams/DecComp1.tikz}{}{\input{./figures/Diagrams/DecComp1.tikz}}\ \ = \ \ %
\InputIfFileExists{Diagrams/DecComp2.tikz}{}{\input{./figures/Diagrams/DecComp2.tikz}}\eeq
Now, using the result we just derived \eqref{eq:C2} we can write that
\beq%
\InputIfFileExists{Diagrams/DecComp2.tikz}{}{\input{./figures/Diagrams/DecComp2.tikz}}\ \ = \ \ %
\InputIfFileExists{Diagrams/DecComp11.tikz}{}{\input{./figures/Diagrams/DecComp11.tikz}}\eeq
Using idempotence and the above result, that is \eqref{eq:C2}, a second time we obtain:
\beq%
\InputIfFileExists{Diagrams/DecComp11.tikz}{}{\input{./figures/Diagrams/DecComp11.tikz}}\ \ = \ \ %
\InputIfFileExists{Diagrams/DecComp12.tikz}{}{\input{./figures/Diagrams/DecComp12.tikz}}\eeq
The second constraint (i.e., eq.~\eqref{eq:25}) then gives us that:
\beq%
\InputIfFileExists{Diagrams/DecComp12.tikz}{}{\input{./figures/Diagrams/DecComp12.tikz}}\ \ = \ \ %
\InputIfFileExists{Diagrams/DecComp13.tikz}{}{\input{./figures/Diagrams/DecComp13.tikz}}\eeq
Finally, using the fact that $f$ and $g$ are decohered means that we can conclude that:
\beq%
\InputIfFileExists{Diagrams/DecComp1.tikz}{}{\input{./figures/Diagrams/DecComp1.tikz}}\ \ = \ \ %
\InputIfFileExists{Diagrams/DecComp14.tikz}{}{\input{./figures/Diagrams/DecComp14.tikz}}\eeq
Hence, the composite of $f$ and $g$ is itself a decohered process.
\endproof

\section{Proof of lemma \ref{lem:34}} \label{app:34}
To see that def.~\ref{def:minConstraint} implies def.~\ref{def:CDIO} is simple. Firstly note that $\mathsf{dec} \subseteq \mathcal{P}^{\mathsf{dec}}$ as, due to idempotency of decoherence processes we have that:
\beq%
\InputIfFileExists{Diagrams/wideDecoherence.tikz}{}{\input{./figures/Diagrams/wideDecoherence.tikz}} \ \ = \ \ %
\InputIfFileExists{Diagrams/wideDecoherenceTriple.tikz}{}{\input{./figures/Diagrams/wideDecoherenceTriple.tikz}}\eeq
therefore, under the assumption of def.~\ref{def:minConstraint}, we have that the composite of $g$ with this decoherence process must itself be a decohered process, that is:
\beq%
\InputIfFileExists{Diagrams/CDIOProof0.tikz}{}{\input{./figures/Diagrams/CDIOProof0.tikz}} \ \ = \ \ %
\InputIfFileExists{Diagrams/CDIOProof1.tikz}{}{\input{./figures/Diagrams/CDIOProof1.tikz}} \ \ = \ \ %
\InputIfFileExists{Diagrams/CDIOProof2.tikz}{}{\input{./figures/Diagrams/CDIOProof2.tikz}} \eeq
where the first equality is simply the definition of a decohered process, and the second is from idempotency. Similarly, we have:
\beq%
\InputIfFileExists{Diagrams/CDIOProof01.tikz}{}{\input{./figures/Diagrams/CDIOProof01.tikz}} \ \ = \ \ %
\InputIfFileExists{Diagrams/CDIOProof11.tikz}{}{\input{./figures/Diagrams/CDIOProof11.tikz}} \ \ = \ \ %
\InputIfFileExists{Diagrams/CDIOProof21.tikz}{}{\input{./figures/Diagrams/CDIOProof21.tikz}} \eeq
combing these we obtain
\beq%
\InputIfFileExists{Diagrams/CDIOProof0.tikz}{}{\input{./figures/Diagrams/CDIOProof0.tikz}} \ \ = \ \  %
\InputIfFileExists{Diagrams/CDIOProof01.tikz}{}{\input{./figures/Diagrams/CDIOProof01.tikz}}\eeq
as required to satisfy def.~\ref{def:CDIO}.

Now, consider the converse direction. Assume we have some decohered process $h$, then
\beq%
\InputIfFileExists{Diagrams/CDIOProof02.tikz}{}{\input{./figures/Diagrams/CDIOProof02.tikz}} \ \ = \ \ %
\InputIfFileExists{Diagrams/CDIOProof12.tikz}{}{\input{./figures/Diagrams/CDIOProof12.tikz}} \ \ = \ \ %
\InputIfFileExists{Diagrams/CDIOProof22.tikz}{}{\input{./figures/Diagrams/CDIOProof22.tikz}} \eeq
where the first equality is from the fact that $h$ is decohered and the second from the assumption of def.~\ref{def:CDIO}. This means that the composite process is indeed a decohered process as we required. Similarly, given a decohered process $h'$ we can show that:
\beq%
\InputIfFileExists{Diagrams/CDIOProof03.tikz}{}{\input{./figures/Diagrams/CDIOProof03.tikz}} \ \ = \ \ %
\InputIfFileExists{Diagrams/CDIOProof13.tikz}{}{\input{./figures/Diagrams/CDIOProof13.tikz}} \ \ = \ \ %
\InputIfFileExists{Diagrams/CDIOProof23.tikz}{}{\input{./figures/Diagrams/CDIOProof23.tikz}} \eeq
and hence, the minimal constraint on free processes is satisfied.

\end{document}

%% file: preamble.tex
\usepackage{rotating}
\usepackage{floatpag}
\rotfloatpagestyle{empty}
\renewcommand{\theenumi}{\arabic{enumi}}
\usepackage{hyperref}
\usepackage{dsfont}
\usepackage{amsmath,amsthm,amssymb}
\usepackage{xspace,enumerate,epsfig}
\usepackage{graphicx}
\graphicspath{{.}{./figures/}}
\usepackage{marvosym}

\usepackage{tikzfig}
\usepackage{stmaryrd}
\usepackage{docmute}
\usepackage{keycommand}

\input{defs.tex}

\input{tikzstyles.tex}

\let\olddagger\dagger
\renewcommand{\dagger}{\ensuremath{\olddagger}\xspace}

\theoremstyle{definition}

\newtheorem{lemma}[]{Lemma}

\newtheorem{definition}[]{Definition}

\newtheorem{example}[]{Example}

\newtheorem{example*}[theorem]{Example*}
\newtheorem{examples*}[theorem]{Examples*}

\newtheorem{remark*}[theorem]{Remark*}

\newtheorem{principle}[]{Principle}

\theoremstyle{plain}

\usepackage{color}
\def\bR{\begin{color}{red}}
\def\bB{\begin{color}{blue}}
\def\bM{\begin{color}{magenta}}
\def\bC{\begin{color}{cyan}}
\def\bW{\begin{color}{white}}
\def\bBl{\begin{color}{black}}
\def\bG{\begin{color}{green}}
\def\bY{\begin{color}{yellow}}
\def\e{\end{color}\xspace}
\newcommand{\bit}{\begin{itemize}}
\newcommand{\eit}{\end{itemize}\par\noindent}
\newcommand{\ben}{\begin{enumerate}}
\newcommand{\een}{\end{enumerate}\par\noindent}
\newcommand{\beq}{\begin{equation}}
\newcommand{\eeq}{\end{equation}\par\noindent}
\newcommand{\beqa}{\begin{eqnarray*}}
\newcommand{\eeqa}{\end{eqnarray*}\par\noindent}
\newcommand{\beqn}{\begin{eqnarray}}
\newcommand{\eeqn}{\end{eqnarray}\par\noindent}

\def\jR{\begin{color}{DarkRed}}
\def\jB{\begin{color}{MidnightBlue}}
\def\jM{\begin{color}{magenta}}
\def\jC{\begin{color}{cyan}}
\def\jW{\begin{color}{white}}
\def\jBl{\begin{color}{black}}
\def\jG{\begin{color}{green}}
\def\jY{\begin{color}{yellow}}

\def\cR{\begin{color}{Crimson}}
\def\cB{\begin{color}{cyan}}
\def\cM{\begin{color}{magenta}}
\def\cC{\begin{color}{cyan}}
\def\cW{\begin{color}{white}}
\def\cBl{\begin{color}{black}}
\def\cG{\begin{color}{green}}
\def\cY{\begin{color}{yellow}}


%% file: defs.tex
\newkeycommand{\pointmap}[style=point][1]{\,\tikz{\node[style=\commandkey{style}] (x) at (0,-0.3) {$#1$};\node [style=none] (3) at (0, 1.0) {};\node [style=none] (3) at (0, -1.0) {};\draw (x)--(0,0.7);}\,}

\newkeycommand{\typedkpoint}[style=kpoint,edgestyle=][2]{%
\,\begin{tikzpicture}
  \begin{pgfonlayer}{nodelayer}
    \node [style=none] (0) at (0, 1) {};
    \node [style=\commandkey{style}] (1) at (0, -0.75) {$#2$};
    \node [style=right label] (2) at (0.25, 0.5) {$#1$};
  \end{pgfonlayer}
  \begin{pgfonlayer}{edgelayer}
    \draw [\commandkey{edgestyle}] (1) to (0.center);
  \end{pgfonlayer}
\end{tikzpicture}\,}

\newkeycommand{\typedkpointdag}[style=kpoint adjoint,edgestyle=][2]{%
\,\begin{tikzpicture}
  \begin{pgfonlayer}{nodelayer}
    \node [style=none] (0) at (0, -1) {};
    \node [style=\commandkey{style}] (1) at (0, 0.75) {$#2$};
    \node [style=right label] (2) at (0.25, -0.5) {$#1$};
  \end{pgfonlayer}
  \begin{pgfonlayer}{edgelayer}
    \draw [\commandkey{edgestyle}] (0.center) to (1);
  \end{pgfonlayer}
\end{tikzpicture}\,}

\newcommand{\bistateadj}[1]{\,\begin{tikzpicture}[yshift=-3mm]
  \begin{pgfonlayer}{nodelayer}
    \node [style=kpointadj, minimum width=1 cm, inner sep=2pt] (0) at (0, 0.5) {$\psi$};
    \node [style=none] (1) at (-0.75, 0.25) {};
    \node [style=none] (2) at (0.75, 0.25) {};
    \node [style=none] (3) at (-0.75, -0.5) {};
    \node [style=none] (4) at (0.75, -0.5) {};
  \end{pgfonlayer}
  \begin{pgfonlayer}{edgelayer}
    \draw (1.center) to (3.center);
    \draw (2.center) to (4.center);
  \end{pgfonlayer}
\end{tikzpicture}\,}

\newkeycommand{\pointketbra}[style1=copoint,style2=point][2]{\,%
\begin{tikzpicture}
  \begin{pgfonlayer}{nodelayer}
    \node [style=none] (0) at (0, -1.5) {};
    \node [style=\commandkey{style1}] (1) at (0, -0.75) {$#1$};
    \node [style=\commandkey{style2}] (2) at (0, 0.75) {$#2$};
    \node [style=none] (3) at (0, 1.5) {};
  \end{pgfonlayer}
  \begin{pgfonlayer}{edgelayer}
    \draw (0.center) to (1);
    \draw (2) to (3.center);
  \end{pgfonlayer}
\end{tikzpicture}\,}

\newkeycommand{\twocopointketbra}[style1=copoint,style2=copoint,style3=point][3]{\,%
\begin{tikzpicture}
  \begin{pgfonlayer}{nodelayer}
    \node [style=none] (0) at (-0.75, -1.5) {};
    \node [style=none] (1) at (0.75, -1.5) {};
    \node [style=\commandkey{style1}] (2) at (-0.75, -0.75) {$#1$};
    \node [style=\commandkey{style2}] (3) at (0.75, -0.75) {$#2$};
    \node [style=\commandkey{style3}] (4) at (0, 0.75) {$#3$};
    \node [style=none] (5) at (0, 1.5) {};
  \end{pgfonlayer}
  \begin{pgfonlayer}{edgelayer}
    \draw (0.center) to (2);
    \draw (1.center) to (3);
    \draw (4) to (5.center);
  \end{pgfonlayer}
\end{tikzpicture}\,}

\newkeycommand{\twopointketbra}[style1=copoint,style2=point,style3=point][3]{\,%
\begin{tikzpicture}
  \begin{pgfonlayer}{nodelayer}
    \node [style=none] (0) at (0, -1.5) {};
    \node [style=none] (1) at (-0.75, 1.5) {};
    \node [style=\commandkey{style1}] (2) at (0, -0.75) {$#1$};
    \node [style=\commandkey{style2}] (3) at (-0.75, 0.75) {$#2$};
    \node [style=\commandkey{style3}] (4) at (0.75, 0.75) {$#3$};
    \node [style=none] (5) at (0.75, 1.5) {};
  \end{pgfonlayer}
  \begin{pgfonlayer}{edgelayer}
    \draw (0.center) to (2);
    \draw (1.center) to (3);
    \draw (4) to (5.center);
  \end{pgfonlayer}
\end{tikzpicture}\,}

\newkeycommand{\pointbraket}[style1=point,style2=copoint][2]{\,%
\begin{tikzpicture}
  \begin{pgfonlayer}{nodelayer}
    \node [style=\commandkey{style1}] (0) at (0, -0.5) {$#1$};
    \node [style=\commandkey{style2}] (1) at (0, 0.5) {$#2$};
  \end{pgfonlayer}
  \begin{pgfonlayer}{edgelayer}
    \draw (0) to (1);
  \end{pgfonlayer}
\end{tikzpicture}\,}

\newkeycommand{\kpointbraket}[style1=kpoint,style2=kpoint adjoint][2]{\,%
\begin{tikzpicture}
  \begin{pgfonlayer}{nodelayer}
    \node [style=\commandkey{style1}] (0) at (0, -0.75) {$#1$};
    \node [style=\commandkey{style2}] (1) at (0, 0.75) {$#2$};
  \end{pgfonlayer}
  \begin{pgfonlayer}{edgelayer}
    \draw (0) to (1);
  \end{pgfonlayer}
\end{tikzpicture}\,}

\newkeycommand{\kpointmap}[style1=kpoint,style2=map][2]{\,%
\begin{tikzpicture}
  \begin{pgfonlayer}{nodelayer}
    \node [style=\commandkey{style1}] (0) at (0, -1.1) {$#1$};
    \node [style=\commandkey{style2}] (1) at (0, 0.2) {$#2$};
    \node [style=none] (2) at (0, 1.5) {};
  \end{pgfonlayer}
  \begin{pgfonlayer}{edgelayer}
    \draw (0) to (1);
    \draw (1) to (2.center);
  \end{pgfonlayer}
\end{tikzpicture}%
\,}

\newkeycommand{\sandwichtwo}[style1=point,style2=map,style3=map,style4=copoint][4]{\,%
\begin{tikzpicture}
  \begin{pgfonlayer}{nodelayer}
    \node [style=\commandkey{style2}] (0) at (0, -0.75) {$#2$};
    \node [style=\commandkey{style3}] (1) at (0, 0.75) {$#3$};
    \node [style=\commandkey{style1}] (2) at (0, -2) {$#1$};
    \node [style=\commandkey{style4}] (3) at (0, 2) {$#4$};
  \end{pgfonlayer}
  \begin{pgfonlayer}{edgelayer}
    \draw (2) to (0);
    \draw (0) to (1);
    \draw (1) to (3);
  \end{pgfonlayer}
\end{tikzpicture}\,}

\newkeycommand{\longbraket}[style1=point,style2=copoint][2]{\,%
\begin{tikzpicture}
  \begin{pgfonlayer}{nodelayer}
    \node [style=\commandkey{style1}] (0) at (0, -2) {$#1$};
    \node [style=\commandkey{style2}] (1) at (0, 2) {$#2$};
  \end{pgfonlayer}
  \begin{pgfonlayer}{edgelayer}
    \draw (0) to (1);
  \end{pgfonlayer}
\end{tikzpicture}\,}

\newkeycommand{\onb}[style=point]{\ensuremath{\left\{\tikz{\node[style=\commandkey{style}] (x) at (0,-0.3) {$j$};\draw (x)--(0,0.7);}\right\}}\xspace}

\newkeycommand{\copointmap}[style=copoint][1]{\,\tikz{\node[style=\commandkey{style}] (x) at (0,0.3) {$#1$};\draw (x)--(0,-0.7);}\,}

\newcommand{\bra}[1]{\ensuremath{\left\langle #1 \right|}}
\newcommand{\ket}[1]{\ensuremath{\left|  #1 \right\rangle}}

\newcommand{\ketbra}[2]{\ensuremath{\ket{#1}\!\bra{#2}}}

\tikzstyle{cdiag}=[matrix of math nodes, row sep=3em, column sep=3em, text height=1.5ex, text depth=0.25ex,inner sep=0.5em]
\tikzstyle{arrow above}=[transform canvas={yshift=0.5ex}]
\tikzstyle{arrow below}=[transform canvas={yshift=-0.5ex}]



%% file: tikzstyles.tex
\usetikzlibrary{shapes.multipart}

\tikzstyle{bwSpider}=[
       rectangle split,
       rectangle split parts=2,
       rectangle split part fill={black,white},
 minimum size=3.6 mm, inner sep=-2mm, draw=black,scale=0.5,rounded corners=0.8 mm
       ]
 \tikzstyle{wbSpider}=[
       rectangle split,
       rectangle split parts=2,
       rectangle split part fill={white,black},
 minimum size=3.6 mm, inner sep=-2mm, draw=black,scale=0.5,rounded corners=0.8 mm
       ]
\tikzstyle{qWire}=[line width = 1pt, color=quantumviolet]
\tikzstyle{cWire}=[color=quantumgray,thin]
\tikzstyle{env}=[copoint,regular polygon rotate=0,minimum width=0.2cm, fill=black]

\tikzstyle{probs}=[shape=semicircle,fill=white,draw=black,shape border rotate=180,minimum width=1.2cm]

\tikzstyle{every picture}=[baseline=-0.25em,scale=0.5]
\tikzstyle{dotpic}=[] 
\tikzstyle{diredges}=[every to/.style={diredge}]
\tikzstyle{math matrix}=[matrix of math nodes,left delimiter=(,right delimiter=),inner sep=2pt,column sep=1em,row sep=0.5em,nodes={inner sep=0pt},text height=1.5ex, text depth=0.25ex]

\tikzstyle{inline text}=[text height=1.5ex, text depth=0.25ex,yshift=0.5mm]
\tikzstyle{label}=[font=\footnotesize,text height=1.5ex, text depth=0.25ex,yshift=0.5mm]
\tikzstyle{left label}=[label,anchor=east,xshift=1.5mm]
\tikzstyle{right label}=[label,anchor=west,xshift=-1mm]

\tikzstyle{braceedge}=[decorate,decoration={brace,amplitude=2mm,raise=-1mm}]
\tikzstyle{small braceedge}=[decorate,decoration={brace,amplitude=1mm,raise=-1mm}]

\tikzstyle{doubled}=[line width=1.6pt] 
\tikzstyle{boldedge}=[doubled,shorten <=-0.17mm,shorten >=-0.17mm]
\tikzstyle{boldedgegray}=[doubled,gray,shorten <=-0.17mm,shorten >=-0.17mm]
\tikzstyle{singleedgegray}=[gray]

\tikzstyle{semidoubled}=[line width=1.4pt] 
\tikzstyle{semiboldedgegray}=[semidoubled,gray,shorten <=-0.17mm,shorten >=-0.17mm]

\tikzstyle{boxedge}=[semiboldedgegray]

\tikzstyle{boldedgedashed}=[very thick,dashed,shorten <=-0.17mm,shorten >=-0.17mm]
\tikzstyle{vboldedgedashed}=[doubled,dashed,shorten <=-0.17mm,shorten >=-0.17mm]
\tikzstyle{left hook arrow}=[left hook-latex]
\tikzstyle{right hook arrow}=[right hook-latex]
\tikzstyle{sembracket}=[line width=0.5pt,shorten <=-0.07mm,shorten >=-0.07mm]

\tikzstyle{causal edge}=[->,thick,gray]
\tikzstyle{causal nondir}=[thick,gray]
\tikzstyle{timeline}=[thick,gray, dashed]

\tikzstyle{cedge}=[<->,thick,gray!70!white]

\tikzstyle{empty diagram}=[draw=gray!40!white,dashed,shape=rectangle,minimum width=1cm,minimum height=1cm]
\tikzstyle{empty diagram small}=[draw=gray!50!white,dashed,shape=rectangle,minimum width=0.6cm,minimum height=0.5cm]

\tikzstyle{dot}=[inner sep=0mm,minimum width=2mm,minimum height=2mm,draw,shape=circle]
\tikzstyle{leak}=[white dot, shape=regular polygon, minimum size=3.3 mm, regular polygon sides=3, outer sep=-0.2mm, regular polygon rotate=270]
\tikzstyle{proj}=[draw,fill=white,chamfered rectangle,chamfered rectangle angle=30, minimum width=2mm, minimum height= 1mm,scale=0.5, outer sep=-0.2mm]
\tikzstyle{wide proj}=[draw,fill=white,chamfered rectangle,chamfered rectangle angle=30, minimum width=15mm,minimum height=1mm,scale=0.5, outer sep=-0.2mm]
\tikzstyle{very wide proj}=[draw,fill=white,chamfered rectangle,chamfered rectangle angle=30, minimum width=25mm,minimum height=1mm,scale=0.5, outer sep=-0.2mm]
\tikzstyle{very very wide proj}=[draw,fill=white,chamfered rectangle,chamfered rectangle angle=30, minimum width=35mm,minimum height=1mm,scale=0.5, outer sep=-0.2mm]
\tikzstyle{preleak}=[proj]

\tikzstyle{split proj out}=[regular polygon,regular polygon sides=3,draw,scale=0.75,inner sep=-0.5pt,minimum width=3.3mm,fill=white,regular polygon rotate=180]
\tikzstyle{split proj in}=[regular polygon,regular polygon sides=3,draw,scale=0.75,inner sep=-0.5pt,minimum width=3.3mm,fill=white]
\tikzstyle{Vleak}=[white dot, shape=regular polygon, minimum size=3.3 mm, regular polygon sides=3, outer sep=-0.2mm, regular polygon rotate=90]
\tikzstyle{dleak}=[white dot, line width=1.6pt, shape=regular polygon, minimum size=3.3 mm, regular polygon sides=3, outer sep=-0.2mm, regular polygon rotate=270]

\tikzstyle{Wsquare}=[white dot, shape=regular polygon, rounded corners=0.8 mm, minimum size=3.3 mm, regular polygon sides=3, outer sep=-0.2mm]
\tikzstyle{Wsquareadj}=[white dot, shape=regular polygon, rounded corners=0.8 mm, minimum size=3.3 mm, regular polygon sides=3, outer sep=-0.2mm, regular polygon rotate=180]
\tikzstyle{ddot}=[inner sep=0mm, doubled, minimum width=2.5mm,minimum height=2.5mm,draw,shape=circle]

\tikzstyle{black dot}=[dot,fill=black]
\tikzstyle{white dot}=[dot,fill=white,,text depth=-0.2mm]
\tikzstyle{white Wsquare}=[Wsquare,fill=gray,,text depth=-0.2mm]
\tikzstyle{white Wsquareadj}=[Wsquareadj,fill=white,,text depth=-0.2mm]
\tikzstyle{green dot}=[white dot] 
\tikzstyle{gray dot}=[dot,fill=gray!40!white,,text depth=-0.2mm]
\tikzstyle{red dot}=[gray dot] 

\tikzstyle{black ddot}=[ddot,fill=black]
\tikzstyle{white ddot}=[ddot,fill=white]
\tikzstyle{gray ddot}=[ddot,fill=gray!40!white]

\tikzstyle{gray edge}=[gray!60!white]

\tikzstyle{small dot}=[inner sep=0.5mm,minimum width=0pt,minimum height=0pt,draw,shape=circle]

\tikzstyle{small black dot}=[small dot,fill=black]
\tikzstyle{small white dot}=[small dot,fill=white]
\tikzstyle{small gray dot}=[small dot,fill=gray!40!white]

\tikzstyle{causal dot}=[inner sep=0.4mm,minimum width=0pt,minimum height=0pt,draw=white,shape=circle,fill=gray!40!white]

\tikzstyle{phase dimensions}=[minimum size=5mm,font=\footnotesize,rectangle,rounded corners=2.5mm,inner sep=0.2mm,outer sep=-2mm]
\tikzstyle{dphase dimensions}=[minimum size=5mm,font=\footnotesize,rectangle,rounded corners=2.5mm,inner sep=0.2mm,outer sep=-2mm]

\tikzstyle{white phase dot}=[dot,fill=white,phase dimensions]
\tikzstyle{white phase ddot}=[ddot,fill=white,dphase dimensions]

\tikzstyle{white rect ddot}=[draw=black,fill=white,doubled,minimum size=5mm,font=\footnotesize,rectangle,rounded corners=2.5mm,inner sep=0.2mm]
\tikzstyle{gray rect ddot}=[draw=black,fill=gray!40!white,doubled,minimum size=6mm,font=\footnotesize,rectangle,rounded corners=3mm]

\tikzstyle{gray phase dot}=[dot,fill=gray!40!white,phase dimensions]
\tikzstyle{gray phase ddot}=[ddot,fill=gray!40!white,dphase dimensions]
\tikzstyle{grey phase dot}=[gray phase dot]
\tikzstyle{grey phase ddot}=[gray phase ddot]

\tikzstyle{small phase dimensions}=[minimum size=4mm,font=\tiny,rectangle,rounded corners=2mm,inner sep=0.2mm,outer sep=-2mm]
\tikzstyle{small dphase dimensions}=[minimum size=4mm,font=\tiny,rectangle,rounded corners=2mm,inner sep=0.2mm,outer sep=-2mm]

\tikzstyle{small gray phase dot}=[dot,fill=gray!40!white,small phase dimensions]
\tikzstyle{small gray phase ddot}=[ddot,fill=gray!40!white,small dphase dimensions]

\tikzstyle{small map}=[draw,shape=rectangle,minimum height=4mm,minimum width=4mm,fill=white]

\tikzstyle{cnot}=[fill=white,shape=circle,inner sep=-1.4pt]

\tikzstyle{asym hadamard}=[fill=white,draw,shape=NEbox,inner sep=0.6mm,font=\footnotesize,minimum height=4mm]
\tikzstyle{asym hadamard conj}=[fill=white,draw,shape=NWbox,inner sep=0.6mm,font=\footnotesize,minimum height=4mm]
\tikzstyle{asym hadamard dag}=[fill=white,draw,shape=SEbox,inner sep=0.6mm,font=\footnotesize,minimum height=4mm]

\tikzstyle{hadamard}=[fill=white,draw,inner sep=0.6mm,font=\footnotesize,minimum height=4mm,minimum width=4mm]
\tikzstyle{small hadamard}=[fill=white,draw,inner sep=0.6mm,minimum height=1.5mm,minimum width=1.5mm]
\tikzstyle{small hadamard rotate}=[small hadamard,rotate=45]
\tikzstyle{dhadamard}=[hadamard,doubled]
\tikzstyle{small dhadamard}=[small hadamard,doubled]
\tikzstyle{small dhadamard rotate}=[small hadamard rotate,doubled]
\tikzstyle{antipode}=[white dot,inner sep=0.3mm,font=\footnotesize]

\tikzstyle{scalar}=[diamond,draw,inner sep=0.5pt,font=\small]
\tikzstyle{dscalar}=[diamond,doubled, draw,inner sep=0.5pt,font=\small]

\tikzstyle{small box}=[rectangle,inline text,fill=white,draw,minimum height=5mm,yshift=-0.5mm,minimum width=5mm,font=\small]
\tikzstyle{small gray box}=[small box,fill=gray!30]
\tikzstyle{medium box}=[rectangle,inline text,fill=white,draw,minimum height=5mm,yshift=-0.5mm,minimum width=10mm,font=\small]
\tikzstyle{square box}=[small box] 
\tikzstyle{medium gray box}=[small box,fill=gray!30]
\tikzstyle{semilarge box}=[rectangle,inline text,fill=white,draw,minimum height=5mm,yshift=-0.5mm,minimum width=12.5mm,font=\small]
\tikzstyle{large box}=[rectangle,inline text,fill=white,draw,minimum height=5mm,yshift=-0.5mm,minimum width=15mm,font=\small]
\tikzstyle{large gray box}=[small box,fill=gray!30]

\tikzstyle{Bayes box}=[rectangle,fill=black,draw, minimum height=3mm, minimum width=3mm]

\tikzstyle{gray square point}=[small box,fill=gray!50]

\tikzstyle{dphase box white}=[dhadamard]
\tikzstyle{dphase box gray}=[dhadamard,fill=gray!50!white]
\tikzstyle{phase box white}=[hadamard]
\tikzstyle{phase box gray}=[hadamard,fill=gray!50!white]

\tikzstyle{point}=[regular polygon,regular polygon sides=3,draw,scale=0.75,inner sep=-0.5pt,minimum width=9mm,fill=white,regular polygon rotate=180]
\tikzstyle{point nosep}=[regular polygon,regular polygon sides=3,draw,scale=0.75,inner sep=-2pt,minimum width=9mm,fill=white,regular polygon rotate=180]
\tikzstyle{copoint}=[regular polygon,regular polygon sides=3,draw,scale=0.75,inner sep=-0.5pt,minimum width=9mm,fill=white]
\tikzstyle{dpoint}=[point,doubled]
\tikzstyle{dcopoint}=[copoint,doubled]

\tikzstyle{pointgrow}=[shape=cornerpoint,kpoint common,scale=0.75,inner sep=3pt]
\tikzstyle{pointgrow dag}=[shape=cornercopoint,kpoint common,scale=0.75,inner sep=3pt]

\tikzstyle{wide copoint}=[fill=white,draw,shape=isosceles triangle,shape border rotate=90,isosceles triangle stretches=true,inner sep=0pt,minimum width=1.5cm,minimum height=6.12mm]
\tikzstyle{wide point}=[fill=white,draw,shape=isosceles triangle,shape border rotate=-90,isosceles triangle stretches=true,inner sep=0pt,minimum width=1.5cm,minimum height=6.12mm,yshift=-0.0mm]
\tikzstyle{wide point plus}=[fill=white,draw,shape=isosceles triangle,shape border rotate=-90,isosceles triangle stretches=true,inner sep=0pt,minimum width=1.74cm,minimum height=7mm,yshift=-0.0mm]

\tikzstyle{wide dpoint}=[fill=white,doubled,draw,shape=isosceles triangle,shape border rotate=-90,isosceles triangle stretches=true,inner sep=0pt,minimum width=1.5cm,minimum height=6.12mm,yshift=-0.0mm]

\tikzstyle{tinypoint}=[regular polygon,regular polygon sides=3,draw,scale=0.55,inner sep=-0.15pt,minimum width=6mm,fill=white,regular polygon rotate=180]

\tikzstyle{white point}=[point]
\tikzstyle{white dpoint}=[dpoint]
\tikzstyle{green point}=[white point] 
\tikzstyle{white copoint}=[copoint]
\tikzstyle{gray point}=[point,fill=gray!40!white]
\tikzstyle{gray dpoint}=[gray point,doubled]
\tikzstyle{red point}=[gray point] 
\tikzstyle{gray copoint}=[copoint,fill=gray!40!white]
\tikzstyle{gray dcopoint}=[gray copoint,doubled]

\tikzstyle{white point guide}=[regular polygon,regular polygon sides=3,font=\scriptsize,draw,scale=0.65,inner sep=-0.5pt,minimum width=9mm,fill=white,regular polygon rotate=180]

\tikzstyle{black point}=[point,fill=black,font=\color{white}]
\tikzstyle{black copoint}=[copoint,fill=black,font=\color{white}]

\tikzstyle{tiny gray point}=[tinypoint,fill=gray!40!white]

\tikzstyle{diredge}=[->]
\tikzstyle{ddiredge}=[<->]
\tikzstyle{rdiredge}=[<-]
\tikzstyle{thickdiredge}=[->, very thick]
\tikzstyle{pointer edge}=[->,very thick,gray]
\tikzstyle{pointer edge part}=[very thick,gray]
\tikzstyle{dashed edge}=[dashed]
\tikzstyle{thick dashed edge}=[very thick,dashed]
\tikzstyle{thick gray dashed edge}=[thick dashed edge,gray!40]
\tikzstyle{thick map edge}=[very thick,|->]

\makeatletter
\newcommand{\boxshape}[3]{%
\pgfdeclareshape{#1}{
\inheritsavedanchors[from=rectangle] 
\inheritanchorborder[from=rectangle]
\inheritanchor[from=rectangle]{center}
\inheritanchor[from=rectangle]{north}
\inheritanchor[from=rectangle]{south}
\inheritanchor[from=rectangle]{west}
\inheritanchor[from=rectangle]{east}
\backgroundpath{
\southwest \pgf@xa=\pgf@x \pgf@ya=\pgf@y
\northeast \pgf@xb=\pgf@x \pgf@yb=\pgf@y

\@tempdima=#2
\@tempdimb=#3

\pgfpathmoveto{\pgfpoint{\pgf@xa - 5pt + \@tempdima}{\pgf@ya}}
\pgfpathlineto{\pgfpoint{\pgf@xa - 5pt - \@tempdima}{\pgf@yb}}
\pgfpathlineto{\pgfpoint{\pgf@xb + 5pt + \@tempdimb}{\pgf@yb}}
\pgfpathlineto{\pgfpoint{\pgf@xb + 5pt - \@tempdimb}{\pgf@ya}}
\pgfpathlineto{\pgfpoint{\pgf@xa - 5pt + \@tempdima}{\pgf@ya}}
\pgfpathclose
}
}}

\boxshape{NEbox}{0pt}{3pt}
\boxshape{SEbox}{0pt}{-3pt}
\boxshape{NWbox}{5pt}{0pt}
\boxshape{SWbox}{-5pt}{0pt}
\boxshape{EBox}{-3pt}{3pt}
\boxshape{WBox}{3pt}{-3pt}
\makeatother

\tikzstyle{cloud}=[shape=cloud,draw,minimum width=1.5cm,minimum height=1.5cm]

\tikzstyle{map}=[draw,shape=NEbox,inner sep=1pt,minimum height=4mm,fill=white]
\tikzstyle{dashedmap}=[draw,dashed,shape=NEbox,inner sep=2pt,minimum height=6mm,fill=white]
\tikzstyle{mapdag}=[draw,shape=SEbox,inner sep=1pt,minimum height=4mm,fill=white]
\tikzstyle{mapadj}=[draw,shape=SEbox,inner sep=2pt,minimum height=6mm,fill=white]
\tikzstyle{maptrans}=[draw,shape=SWbox,inner sep=2pt,minimum height=6mm,fill=white]
\tikzstyle{mapconj}=[draw,shape=NWbox,inner sep=2pt,minimum height=6mm,fill=white]

\tikzstyle{medium map}=[draw,shape=NEbox,inner sep=2pt,minimum height=6mm,fill=white,minimum width=7mm]
\tikzstyle{medium map dag}=[draw,shape=SEbox,inner sep=2pt,minimum height=6mm,fill=white,minimum width=7mm]
\tikzstyle{medium map adj}=[draw,shape=SEbox,inner sep=2pt,minimum height=6mm,fill=white,minimum width=7mm]
\tikzstyle{medium map trans}=[draw,shape=SWbox,inner sep=2pt,minimum height=6mm,fill=white,minimum width=7mm]
\tikzstyle{medium map conj}=[draw,shape=NWbox,inner sep=2pt,minimum height=6mm,fill=white,minimum width=7mm]
\tikzstyle{semilarge map}=[draw,shape=NEbox,inner sep=2pt,minimum height=6mm,fill=white,minimum width=9.5mm]
\tikzstyle{semilarge map trans}=[draw,shape=SWbox,inner sep=2pt,minimum height=6mm,fill=white,minimum width=9.5mm]
\tikzstyle{semilarge map adj}=[draw,shape=SEbox,inner sep=2pt,minimum height=6mm,fill=white,minimum width=9.5mm]
\tikzstyle{semilarge map dag}=[draw,shape=SEbox,inner sep=2pt,minimum height=6mm,fill=white,minimum width=9.5mm]
\tikzstyle{semilarge map conj}=[draw,shape=NWbox,inner sep=2pt,minimum height=6mm,fill=white,minimum width=9.5mm]
\tikzstyle{large map}=[draw,shape=NEbox,inner sep=2pt,minimum height=6mm,fill=white,minimum width=12mm]
\tikzstyle{large map conj}=[draw,shape=NWbox,inner sep=2pt,minimum height=6mm,fill=white,minimum width=12mm]
\tikzstyle{very large map}=[draw,shape=NEbox,inner sep=2pt,minimum height=6mm,fill=white,minimum width=17mm]

\tikzstyle{medium dmap}=[draw,doubled,shape=NEbox,inner sep=2pt,minimum height=6mm,fill=white,minimum width=7mm]
\tikzstyle{medium dmap dag}=[draw,doubled,shape=SEbox,inner sep=2pt,minimum height=6mm,fill=white,minimum width=7mm]
\tikzstyle{medium dmap adj}=[draw,doubled,shape=SEbox,inner sep=2pt,minimum height=6mm,fill=white,minimum width=7mm]
\tikzstyle{medium dmap trans}=[draw,doubled,shape=SWbox,inner sep=2pt,minimum height=6mm,fill=white,minimum width=7mm]
\tikzstyle{medium dmap conj}=[draw,doubled,shape=NWbox,inner sep=2pt,minimum height=6mm,fill=white,minimum width=7mm]
\tikzstyle{semilarge dmap}=[draw,doubled,shape=NEbox,inner sep=2pt,minimum height=6mm,fill=white,minimum width=9.5mm]
\tikzstyle{semilarge dmap trans}=[draw,doubled,shape=SWbox,inner sep=2pt,minimum height=6mm,fill=white,minimum width=9.5mm]
\tikzstyle{semilarge dmap adj}=[draw,doubled,shape=SEbox,inner sep=2pt,minimum height=6mm,fill=white,minimum width=9.5mm]
\tikzstyle{semilarge dmap dag}=[draw,doubled,shape=SEbox,inner sep=2pt,minimum height=6mm,fill=white,minimum width=9.5mm]
\tikzstyle{semilarge dmap conj}=[draw,doubled,shape=NWbox,inner sep=2pt,minimum height=6mm,fill=white,minimum width=9.5mm]
\tikzstyle{large dmap}=[draw,doubled,shape=NEbox,inner sep=2pt,minimum height=6mm,fill=white,minimum width=12mm]
\tikzstyle{large dmap conj}=[draw,doubled,shape=NWbox,inner sep=2pt,minimum height=6mm,fill=white,minimum width=12mm]
\tikzstyle{large dmap trans}=[draw,doubled,shape=SWbox,inner sep=2pt,minimum height=6mm,fill=white,minimum width=12mm]
\tikzstyle{large dmap adj}=[draw,doubled,shape=SEbox,inner sep=2pt,minimum height=6mm,fill=white,minimum width=12mm]
\tikzstyle{large dmap dag}=[draw,doubled,shape=SEbox,inner sep=2pt,minimum height=6mm,fill=white,minimum width=12mm]
\tikzstyle{very large dmap}=[draw,doubled,shape=NEbox,inner sep=2pt,minimum height=6mm,fill=white,minimum width=19.5mm]

\tikzstyle{muxbox}=[draw,shape=rectangle,minimum height=3mm,minimum width=3mm,fill=white]
\tikzstyle{dmuxbox}=[muxbox,doubled]

\tikzstyle{box}=[draw,shape=rectangle,inner sep=2pt,minimum height=6mm,minimum width=6mm,fill=white]
\tikzstyle{dbox}=[draw,doubled,shape=rectangle,inner sep=2pt,minimum height=6mm,minimum width=6mm,fill=white]
\tikzstyle{dmap}=[draw,doubled,shape=NEbox,inner sep=2pt,minimum height=6mm,fill=white]
\tikzstyle{dmapdag}=[draw,doubled,shape=SEbox,inner sep=2pt,minimum height=6mm,fill=white]
\tikzstyle{dmapadj}=[draw,doubled,shape=SEbox,inner sep=2pt,minimum height=6mm,fill=white]
\tikzstyle{dmaptrans}=[draw,doubled,shape=SWbox,inner sep=2pt,minimum height=6mm,fill=white]
\tikzstyle{dmapconj}=[draw,doubled,shape=NWbox,inner sep=2pt,minimum height=6mm,fill=white]

\tikzstyle{ddmap}=[draw,doubled,dashed,shape=NEbox,inner sep=2pt,minimum height=6mm,fill=white]
\tikzstyle{ddmapdag}=[draw,doubled,dashed,shape=SEbox,inner sep=2pt,minimum height=6mm,fill=white]
\tikzstyle{ddmapadj}=[draw,doubled,dashed,shape=SEbox,inner sep=2pt,minimum height=6mm,fill=white]
\tikzstyle{ddmaptrans}=[draw,doubled,dashed,shape=SWbox,inner sep=2pt,minimum height=6mm,fill=white]
\tikzstyle{ddmapconj}=[draw,doubled,dashed,shape=NWbox,inner sep=2pt,minimum height=6mm,fill=white]

\boxshape{sNEbox}{0pt}{3pt}
\boxshape{sSEbox}{0pt}{-3pt}
\boxshape{sNWbox}{3pt}{0pt}
\boxshape{sSWbox}{-3pt}{0pt}
\tikzstyle{smap}=[draw,shape=sNEbox,fill=white]
\tikzstyle{smapdag}=[draw,shape=sSEbox,fill=white]
\tikzstyle{smapadj}=[draw,shape=sSEbox,fill=white]
\tikzstyle{smaptrans}=[draw,shape=sSWbox,fill=white]
\tikzstyle{smapconj}=[draw,shape=sNWbox,fill=white]

\tikzstyle{dsmap}=[draw,dashed,shape=sNEbox,fill=white]
\tikzstyle{dsmapdag}=[draw,dashed,shape=sSEbox,fill=white]
\tikzstyle{dsmaptrans}=[draw,dashed,shape=sSWbox,fill=white]
\tikzstyle{dsmapconj}=[draw,dashed,shape=sNWbox,fill=white]

\boxshape{mNEbox}{0pt}{10pt}
\boxshape{mSEbox}{0pt}{-10pt}
\boxshape{mNWbox}{10pt}{0pt}
\boxshape{mSWbox}{-10pt}{0pt}
\tikzstyle{mmap}=[draw,shape=mNEbox]
\tikzstyle{mmapdag}=[draw,shape=mSEbox]
\tikzstyle{mmaptrans}=[draw,shape=mSWbox]
\tikzstyle{mmapconj}=[draw,shape=mNWbox]

\tikzstyle{mmapgray}=[draw,fill=gray!40!white,shape=mNEbox]
\tikzstyle{smapgray}=[draw,fill=gray!40!white,shape=sNEbox]

\makeatletter

\pgfdeclareshape{cornerpoint}{
\inheritsavedanchors[from=rectangle] 
\inheritanchorborder[from=rectangle]
\inheritanchor[from=rectangle]{center}
\inheritanchor[from=rectangle]{north}
\inheritanchor[from=rectangle]{south}
\inheritanchor[from=rectangle]{west}
\inheritanchor[from=rectangle]{east}
\backgroundpath{
\southwest \pgf@xa=\pgf@x \pgf@ya=\pgf@y
\northeast \pgf@xb=\pgf@x \pgf@yb=\pgf@y

\pgfmathsetmacro{\pgf@shorten@left}{\pgfkeysvalueof{/tikz/shorten left}}
\pgfmathsetmacro{\pgf@shorten@right}{\pgfkeysvalueof{/tikz/shorten right}}

\pgfpathmoveto{\pgfpoint{0.5 * (\pgf@xa + \pgf@xb)}{\pgf@ya - 5pt}}
\pgfpathlineto{\pgfpoint{\pgf@xa - 8pt + \pgf@shorten@left}{\pgf@yb - 1.5 * \pgf@shorten@left}}
\pgfpathlineto{\pgfpoint{\pgf@xa - 8pt + \pgf@shorten@left}{\pgf@yb}}
\pgfpathlineto{\pgfpoint{\pgf@xb + 8pt - \pgf@shorten@right}{\pgf@yb}}
\pgfpathlineto{\pgfpoint{\pgf@xb + 8pt - \pgf@shorten@right}{\pgf@yb - 1.5 * \pgf@shorten@right}}
\pgfpathclose
}
}

\pgfdeclareshape{cornercopoint}{
\inheritsavedanchors[from=rectangle] 
\inheritanchorborder[from=rectangle]
\inheritanchor[from=rectangle]{center}
\inheritanchor[from=rectangle]{north}
\inheritanchor[from=rectangle]{south}
\inheritanchor[from=rectangle]{west}
\inheritanchor[from=rectangle]{east}
\backgroundpath{
\southwest \pgf@xa=\pgf@x \pgf@ya=\pgf@y
\northeast \pgf@xb=\pgf@x \pgf@yb=\pgf@y

\pgfmathsetmacro{\pgf@shorten@left}{\pgfkeysvalueof{/tikz/shorten left}}
\pgfmathsetmacro{\pgf@shorten@right}{\pgfkeysvalueof{/tikz/shorten right}}

\pgfpathmoveto{\pgfpoint{0.5 * (\pgf@xa + \pgf@xb)}{\pgf@yb + 5pt}}
\pgfpathlineto{\pgfpoint{\pgf@xa - 8pt + \pgf@shorten@left}{\pgf@ya + 1.5 * \pgf@shorten@left}}
\pgfpathlineto{\pgfpoint{\pgf@xa - 8pt + \pgf@shorten@left}{\pgf@ya}}
\pgfpathlineto{\pgfpoint{\pgf@xb + 8pt - \pgf@shorten@right}{\pgf@ya}}
\pgfpathlineto{\pgfpoint{\pgf@xb + 8pt - \pgf@shorten@right}{\pgf@ya + 1.5 * \pgf@shorten@right}}
\pgfpathclose
}
}

\makeatother

\pgfkeyssetvalue{/tikz/shorten left}{0pt}
\pgfkeyssetvalue{/tikz/shorten right}{0pt}

\tikzstyle{kpoint common}=[draw,fill=white,inner sep=1pt,minimum height=4mm]
\tikzstyle{kpoint sc}=[shape=cornerpoint,kpoint common]
\tikzstyle{kpoint adjoint sc}=[shape=cornercopoint,kpoint common]
\tikzstyle{kpoint}=[shape=cornerpoint,shorten left=5pt,kpoint common]
\tikzstyle{kpoint adjoint}=[shape=cornercopoint,shorten left=5pt,kpoint common]
\tikzstyle{kpoint conjugate}=[shape=cornerpoint,shorten right=5pt,kpoint common]
\tikzstyle{kpoint transpose}=[shape=cornercopoint,shorten right=5pt,kpoint common]
\tikzstyle{kpoint symm}=[shape=cornerpoint,shorten left=5pt,shorten right=5pt,kpoint common]

\tikzstyle{wide kpoint sc}=[shape=cornerpoint,kpoint common, minimum width=1 cm]
\tikzstyle{wide kpointdag sc}=[shape=cornercopoint,kpoint common, minimum width=1 cm]

\tikzstyle{black kpoint}=[shape=cornerpoint,shorten left=5pt,kpoint common,fill=black,font=\color{white}]

\tikzstyle{black kpoint sm}=[shape=cornerpoint,shorten left=5pt,kpoint common,fill=black,font=\color{white},scale=0.75]

\tikzstyle{black kpoint adjoint}=[shape=cornercopoint,shorten left=5pt,kpoint common,fill=black,font=\color{white}]
\tikzstyle{black kpointadj}=[shape=cornercopoint,shorten left=5pt,kpoint common,fill=black,font=\color{white}]

\tikzstyle{black kpointadj sm}=[shape=cornercopoint,shorten left=5pt,kpoint common,fill=black,font=\color{white},scale=0.75]

\tikzstyle{black dkpoint}=[shape=cornerpoint,shorten left=5pt,kpoint common,fill=black, doubled,font=\color{white}]
\tikzstyle{black dkpoint adjoint}=[shape=cornercopoint,shorten left=5pt,kpoint common,fill=black, doubled,font=\color{white}]
\tikzstyle{black dkpointadj}=[shape=cornercopoint,shorten left=5pt,kpoint common,fill=black, doubled,font=\color{white}]

\tikzstyle{black dkpoint sm}=[shape=cornerpoint,shorten left=5pt,kpoint common,fill=black, doubled,font=\color{white},scale=0.75]
\tikzstyle{black dkpointadj sm}=[shape=cornercopoint,shorten left=5pt,kpoint common,fill=black, doubled,font=\color{white},scale=0.75]

\tikzstyle{kpointdag}=[kpoint adjoint]
\tikzstyle{kpointadj}=[kpoint adjoint]
\tikzstyle{kpointconj}=[kpoint conjugate]
\tikzstyle{kpointtrans}=[kpoint transpose]

\tikzstyle{big kpoint}=[kpoint, minimum width=1.2 cm, minimum height=8mm, inner sep=4pt, text depth=3mm]

\tikzstyle{wide kpoint}=[kpoint, minimum width=1 cm, inner sep=2pt]
\tikzstyle{wide kpointdag}=[kpointdag, minimum width=1 cm, inner sep=2pt]
\tikzstyle{wide kpointconj}=[kpointconj, minimum width=1 cm, inner sep=2pt]
\tikzstyle{wide kpointtrans}=[kpointtrans, minimum width=1 cm, inner sep=2pt]

\tikzstyle{wider kpoint}=[kpoint, minimum width=1.25 cm, inner sep=2pt]
\tikzstyle{wider kpointdag}=[kpointdag, minimum width=1.25 cm, inner sep=2pt]
\tikzstyle{wider kpointconj}=[kpointconj, minimum width=1.25 cm, inner sep=2pt]
\tikzstyle{wider kpointtrans}=[kpointtrans, minimum width=1.25 cm, inner sep=2pt]

\tikzstyle{gray kpoint}=[kpoint,fill=gray!50!white]
\tikzstyle{gray kpointdag}=[kpointdag,fill=gray!50!white]
\tikzstyle{gray kpointadj}=[kpointadj,fill=gray!50!white]
\tikzstyle{gray kpointconj}=[kpointconj,fill=gray!50!white]
\tikzstyle{gray kpointtrans}=[kpointtrans,fill=gray!50!white]

\tikzstyle{gray dkpoint}=[kpoint,fill=gray!50!white,doubled]
\tikzstyle{gray dkpointdag}=[kpointdag,fill=gray!50!white,doubled]
\tikzstyle{gray dkpointadj}=[kpointadj,fill=gray!50!white,doubled]
\tikzstyle{gray dkpointconj}=[kpointconj,fill=gray!50!white,doubled]
\tikzstyle{gray dkpointtrans}=[kpointtrans,fill=gray!50!white,doubled]

\tikzstyle{white label}=[draw,fill=white,rectangle,inner sep=0.7 mm]
\tikzstyle{gray label}=[draw,fill=gray!50!white,rectangle,inner sep=0.7 mm]
\tikzstyle{black label}=[draw,fill=black,rectangle,inner sep=0.7 mm]

\tikzstyle{dkpoint}=[kpoint,doubled]
\tikzstyle{wide dkpoint}=[wide kpoint,doubled]
\tikzstyle{dkpointdag}=[kpoint adjoint,doubled]
\tikzstyle{wide dkpointdag}=[wide kpointdag,doubled]
\tikzstyle{dkcopoint}=[kpoint adjoint,doubled]
\tikzstyle{dkpointadj}=[kpoint adjoint,doubled]
\tikzstyle{dkpointconj}=[kpoint conjugate,doubled]
\tikzstyle{dkpointtrans}=[kpoint transpose,doubled]

\tikzstyle{kscalar}=[kpoint common, shape=EBox, inner xsep=-1pt, inner ysep=3pt,font=\small]
\tikzstyle{kscalarconj}=[kpoint common, shape=WBox, inner xsep=-1pt, inner ysep=3pt,font=\small]

\tikzstyle{spekpoint}=[kpoint sc,minimum height=5mm,inner sep=3pt]
\tikzstyle{spekcopoint}=[kpoint adjoint sc,minimum height=5mm,inner sep=3pt]

\tikzstyle{dspekpoint}=[spekpoint,doubled]
\tikzstyle{dspekcopoint}=[spekcopoint,doubled]

 \tikzstyle{upground}=[circuit ee IEC,thick,ground,rotate=90,scale=2.5]
 \tikzstyle{downground}=[circuit ee IEC,thick,ground,rotate=-90,scale=2.5]
 \tikzstyle{bigground}=[regular polygon,regular polygon sides=3,draw=gray,scale=0.50,inner sep=-0.5pt,minimum width=10mm,fill=gray]

\tikzstyle{arrs}=[-latex,font=\small,auto]
\tikzstyle{arrow plain}=[arrs]
\tikzstyle{arrow dashed}=[dashed,arrs]
\tikzstyle{arrow bold}=[very thick,arrs]
\tikzstyle{arrow hide}=[draw=white!0,-]
\tikzstyle{arrow reverse}=[latex-]
\tikzstyle{cdnode}=[]

%% file: Diagrams/projWhite.tikz
\begin{tikzpicture}
	\begin{pgfonlayer}{nodelayer}
		\node [style=proj] (1000) at (0, -0) {};
	\end{pgfonlayer}
	\begin{pgfonlayer}{edgelayer}
	\end{pgfonlayer}
\end{tikzpicture}

%% file: Diagrams/projBlack.tikz
\begin{tikzpicture}
	\begin{pgfonlayer}{nodelayer}
		\node [style=proj,fill=black] (1000) at (0, -0) {};
	\end{pgfonlayer}
	\begin{pgfonlayer}{edgelayer}
	\end{pgfonlayer}
\end{tikzpicture}

%% file: Diagrams/discardA.tikz
\begin{tikzpicture}
	\begin{pgfonlayer}{nodelayer}
		\node [style=none] (0) at (0, -0) {};
		\node [style=none] (1) at (0, -0.75) {};
		\node [style=upground] (2) at (0, 0.25) {};
		\node [style={right label}] (3) at (0, -0.75) {$A$};
	\end{pgfonlayer}
	\begin{pgfonlayer}{edgelayer}
		\draw[qWire] (0.center) to (1.center);
	\end{pgfonlayer}
\end{tikzpicture}

%% file: Diagrams/discardAB.tikz
\begin{tikzpicture}
	\begin{pgfonlayer}{nodelayer}
		\node [style=none] (0) at (0, -0) {};
		\node [style=none] (1) at (0, -0.75) {};
		\node [style=upground] (2) at (0, 0.25) {};
		\node [style={right label}] (3) at (0, -0.75) {$A B$};
	\end{pgfonlayer}
	\begin{pgfonlayer}{edgelayer}
		\draw[qWire] (0.center) to (1.center);
	\end{pgfonlayer}
\end{tikzpicture}

%% file: Diagrams/discardB.tikz
\begin{tikzpicture}
	\begin{pgfonlayer}{nodelayer}
		\node [style=none] (0) at (0, -0) {};
		\node [style=none] (1) at (0, -0.75) {};
		\node [style=upground] (2) at (0, 0.25) {};
		\node [style={right label}] (3) at (0, -0.75) {$B$};
	\end{pgfonlayer}
	\begin{pgfonlayer}{edgelayer}
		\draw[qWire] (0.center) to (1.center);
	\end{pgfonlayer}
\end{tikzpicture}

%% file: Diagrams/identityA.tikz
\begin{tikzpicture}
	\begin{pgfonlayer}{nodelayer}
		\node [style={right label}] (0) at (0, -0) {$A$};
		\node [style=none] (1) at (0, 1) {};
		\node [style=none] (2) at (0, -1) {};
	\end{pgfonlayer}
	\begin{pgfonlayer}{edgelayer}
		\draw [qWire] (1.center) to (2.center);
	\end{pgfonlayer}
\end{tikzpicture}

%% file: Diagrams/MinConstraint1.tikz
\begin{tikzpicture}
	\begin{pgfonlayer}{nodelayer}
		\node [style={small box}] (0) at (0, -0) {$g$};
		\node [style=proj] (1) at (0, -1) {};
		\node [style=point] (2) at (0, -2) {$s$};
		\node [style=none] (3) at (0, 1) {};
	\end{pgfonlayer}
	\begin{pgfonlayer}{edgelayer}
		\draw [qWire] (2) to (1);
		\draw [qWire] (1) to (0);
		\draw [qWire] (0) to (3.center);
	\end{pgfonlayer}
\end{tikzpicture}

%% file: Diagrams/SimpleRTS2.tikz
\begin{tikzpicture}
	\begin{pgfonlayer}{nodelayer}
		\node [style={small box}] (0) at (0, -0) {$\mathcal{E}$};
		\node [style=point] (1) at (0, -1) {$\rho$};
		\node [style=none] (2) at (0, 1) {};
	\end{pgfonlayer}
	\begin{pgfonlayer}{edgelayer}
		\draw [qWire] (1) to (0);
		\draw [qWire] (0) to (2.center);
	\end{pgfonlayer}
\end{tikzpicture}

%% file: Diagrams/SimpleRTS3.tikz
\begin{tikzpicture}
	\begin{pgfonlayer}{nodelayer}
		\node [style=none] (0) at (0, 1) {};
		\node [style=point] (1) at (0, -0) {$\sigma$};
	\end{pgfonlayer}
	\begin{pgfonlayer}{edgelayer}
		\draw [qWire] (1) to (0.center);
	\end{pgfonlayer}
\end{tikzpicture}

%% file: Diagrams/orthogStates.tikz
\begin{tikzpicture}
	\begin{pgfonlayer}{nodelayer}
		\node [style=point] (0) at (0, -0) {$s_\gamma$};
		\node [style=none] (1) at (0, 1) {};
	\end{pgfonlayer}
	\begin{pgfonlayer}{edgelayer}
		\draw [qWire] (1.center) to (0);
	\end{pgfonlayer}
\end{tikzpicture}

%% file: Diagrams/orthogEffects.tikz
\begin{tikzpicture}
	\begin{pgfonlayer}{nodelayer}
		\node [style=copoint] (0) at (0, 0.25) {$e_\gamma$};
		\node [style=none] (1) at (0, -0.75) {};
	\end{pgfonlayer}
	\begin{pgfonlayer}{edgelayer}
		\draw [qWire] (1.center) to (0);
	\end{pgfonlayer}
\end{tikzpicture}

%% file: Diagrams/groupAction.tikz
\begin{tikzpicture}
	\begin{pgfonlayer}{nodelayer}
		\node [style={small box}] (0) at (0, -0) {$R_\gamma$};
		\node [style=none] (1) at (0, 1) {};
		\node [style=none] (2) at (0, -1) {};
	\end{pgfonlayer}
	\begin{pgfonlayer}{edgelayer}
		\draw [qWire] (1.center) to (0);
		\draw [qWire] (0) to (2.center);
	\end{pgfonlayer}
\end{tikzpicture}

%% file: Diagrams/Karoubi2.tikz
\begin{tikzpicture}
	\begin{pgfonlayer}{nodelayer}
		\node [style=none] (0) at (0, -0) {};
		\node [style=none] (1) at (0, -1) {};
		\node [style=none] (2) at (0, 1) {};
	\end{pgfonlayer}
	\begin{pgfonlayer}{edgelayer}
		\draw [qWire] (2.center) to (0.center);
		\draw [qWire] (0.center) to (1.center);
	\end{pgfonlayer}
\end{tikzpicture}

%% file: Diagrams/smallDecohere.tikz
\begin{tikzpicture}
	\begin{pgfonlayer}{nodelayer}
		\node [style=proj] (0) at (0, -0) {};
		\node [style=none] (1) at (0, -0.5) {};
		\node [style=none] (2) at (0, 0.5) {};
	\end{pgfonlayer}
	\begin{pgfonlayer}{edgelayer}
		\draw [qWire] (2.center) to (0);
		\draw [qWire] (0) to (1.center);
	\end{pgfonlayer}
\end{tikzpicture}

%% file: Diagrams/RTofStates.tikz
\begin{tikzpicture}
	\begin{pgfonlayer}{nodelayer}
		\node [style=none] (0) at (0, -1) {};
		\node [style=none] (1) at (0, 1) {};
		\node [style={right label}] (2) at (0, -0) {$s^A$};
	\end{pgfonlayer}
	\begin{pgfonlayer}{edgelayer}
		\draw (0.center) to (1.center);
	\end{pgfonlayer}
\end{tikzpicture}

%% file: Diagrams/RTofStates2.tikz
\begin{tikzpicture}
	\begin{pgfonlayer}{nodelayer}
		\node [style=point] (0) at (0, -0.25) {$s$};
		\node [style=none] (1) at (0, 1) {};
		\node [style={right label}] (2) at (0, 0.75) {$A$};
	\end{pgfonlayer}
	\begin{pgfonlayer}{edgelayer}
		\draw [qWire] (0) to (1.center);
	\end{pgfonlayer}
\end{tikzpicture}

%% file: Diagrams/RTofStates6.tikz
\begin{tikzpicture}
	\begin{pgfonlayer}{nodelayer}
		\node [style=point] (0) at (0, -0) {$r$};
		\node [style=none] (1) at (0, 1.25) {};
		\node [style={right label}] (2) at (0, 1) {$B$};
	\end{pgfonlayer}
	\begin{pgfonlayer}{edgelayer}
		\draw [qWire] (0) to (1.center);
	\end{pgfonlayer}
\end{tikzpicture}

%% file: Diagrams/RTofMeasurements.tikz
\begin{tikzpicture}
	\begin{pgfonlayer}{nodelayer}
		\node [style=none] (0) at (0, -1) {};
		\node [style=none] (1) at (0, 1) {};
		\node [style={right label}] (2) at (0, -0) {$M_A^n$};
	\end{pgfonlayer}
	\begin{pgfonlayer}{edgelayer}
		\draw (0.center) to (1.center);
	\end{pgfonlayer}
\end{tikzpicture}

%% file: Diagrams/RTofProcesses.tikz
\begin{tikzpicture}
	\begin{pgfonlayer}{nodelayer}
		\node [style=none] (0) at (0, -1) {};
		\node [style=none] (1) at (0, 1) {};
		\node [style={right label}] (2) at (0, -0) {$f^B_A$};
	\end{pgfonlayer}
	\begin{pgfonlayer}{edgelayer}
		\draw (0.center) to (1.center);
	\end{pgfonlayer}
\end{tikzpicture}